\documentclass[twocolumn,amsthm]{autart}

\newcommand{\mypar}{\\[0.2cm]}

\usepackage{mathtools,amsthm,amssymb,amsmath,amsfonts,bbm}
\usepackage[mathscr]{euscript}

\usepackage[bb=dsserif]{mathalpha}

\usepackage{graphicx,psfrag}
\usepackage{xcolor}
\usepackage{float}
\usepackage{multicol}
\usepackage{tcolorbox}
\usepackage{makecell}

\usepackage{cite}

 \usepackage{subfigure}
 \usepackage[english]{babel}
 \usepackage[hidelinks]{hyperref}
 \usepackage{dsfont}

 \usepackage[shortlabels]{enumitem}
 
 \usepackage{epsfig}

\renewcommand{\qed}{\hfill\blacksquare}
\newcommand{\qedwhite}{\hfill \square}

\theoremstyle{plain}
\newtheorem{theorem}{Theorem}
\newtheorem{lemma}{Lemma}
\newtheorem{corollary}{Corollary}
\newtheorem{proposition}{Proposition}

\theoremstyle{definition}
\newtheorem{example}{Example}
\newtheorem{assumption}{Assumption}
\newtheorem{sassumption}{Standing Assumption}
\newtheorem{definition}{Definition}

\newtheorem{problem}{Problem}

\theoremstyle{remark}
\newtheorem{remark}{Remark}

\DeclareMathOperator{\col}{col}
\DeclareMathOperator{\im}{im}

\DeclareMathOperator{\bdiag}{blockdiag}

\DeclareMathOperator{\rank}{rank}

\DeclareMathOperator{\spec}{spec}

\newcommand{\norm}[1]{\ensuremath{\left\| #1 \right\|}}

\DeclareMathOperator{\mspec}{mspec}

\newcommand{\calA}{\ensuremath{\mathcal{A}}}

\newcommand{\calC}{\ensuremath{\mathcal{C}}}

\newcommand{\calF}{\ensuremath{\mathcal{F}}}
\newcommand{\calG}{\ensuremath{\mathcal{G}}}
\newcommand{\calI}{\ensuremath{\mathcal{I}}}
\newcommand{\calJ}{\ensuremath{\mathcal{J}}}

\newcommand{\calO}{\ensuremath{\mathcal{O}}}

\newcommand{\calQ}{\ensuremath{\mathcal{Q}}}
\newcommand{\calR}{\ensuremath{\mathcal{R}}}
\newcommand{\calS}{\ensuremath{\mathcal{S}}}
\newcommand{\calT}{\ensuremath{\mathcal{T}}}

\newcommand{\bT}{\ensuremath{\boldsymbol{T}}}
\newcommand{\bF}{\ensuremath{\boldsymbol{F}}}
\newcommand{\bG}{\ensuremath{\boldsymbol{G}}}
\newcommand{\bS}{\ensuremath{\boldsymbol{S}}}

\newcommand{\bP}{\ensuremath{\boldsymbol{P}}}
\newcommand{\bL}{\ensuremath{\boldsymbol{L}}}
\newcommand{\bro}{\ensuremath{\boldsymbol{r_1}}}
\newcommand{\brt}{\ensuremath{\boldsymbol{r_2}}}
\newcommand{\brth}{\ensuremath{\boldsymbol{r_3}}}
\newcommand{\bR}{\ensuremath{\boldsymbol{R}}}
\newcommand{\br}{\ensuremath{\boldsymbol{r}}}

\newcommand{\bbf}{\ensuremath{\boldsymbol{f}}}

\newcommand{\cbT}{\ensuremath{\boldsymbol{\bar{T}}}}
\newcommand{\cbF}{\ensuremath{\boldsymbol{\bar{F}}}}
\newcommand{\cbG}{\ensuremath{\boldsymbol{\bar{G}}}}
\newcommand{\cbS}{\ensuremath{\boldsymbol{\bar{S}}}}
\newcommand{\cbQ}{\ensuremath{\boldsymbol{Q}}}
\newcommand{\cbb}{\ensuremath{\boldsymbol{b}}}
\newcommand{\bY}{\ensuremath{\bar{Y}}}

\newcommand{\bx}{\ensuremath{\boldsymbol{x}}}
\newcommand{\bu}{\ensuremath{\boldsymbol{u}}}
\newcommand{\bxt}{\ensuremath{\boldsymbol{\tilde{x}}}}
\newcommand{\but}{\ensuremath{\boldsymbol{\tilde{u}}}}
\newcommand{\bo}{\ensuremath{\mathbb{1}}}

\newcommand{\R}{\ensuremath{\mathbb R}}
\newcommand{\N}{\ensuremath{\mathbb N}}

\newcommand{\RNum}[1]{\uppercase\expandafter{\romannumeral #1\relax}}

\usepackage[prependcaption,colorinlistoftodos]{todonotes}

\newcommand{\bse}{\begin{subequations}}
	\newcommand{\ese}{\end{subequations}}
\def\be{\begin{equation}}
\def\ee{\end{equation}}

\newcommand{\bbm}{\begin{bmatrix}}
	\newcommand{\ebm}{\end{bmatrix}}

\usepackage[normalem]{ulem}

\definecolor{ao(english)}{rgb}{0.0, 1, 0.0}

\usepackage{url}
\usepackage[hidelinks]{hyperref}
\newcommand{\oprocendsymbol}{\hbox{$\bullet$}}
\newcommand{\oprocend}{\relax\ifmmode\else\unskip\hfill\fi\oprocendsymbol}

\begin{document}

\begin{frontmatter}
\title{  %
{Privacy Analysis of Affine Transformations in Cloud-based MPC: Vulnerability to Side-knowledge}}
\vspace{-0.5cm}
\author[RUG]{Teimour Hosseinalizadeh}\ead{t.hosseinalizadeh@rug.nl},$\,$
\author[TUD]{Nils Schl{\"u}ter}\ead{nils.schlueter@tu-dortmund.de},$\,$
\author[TUD]{Moritz Schulze Darup}\ead{moritz.schulzedarup@tu-dortmund.de},$\,$
\author[RUG]{Nima Monshizadeh}\ead{n.monshizadeh@rug.nl}    %
\address[RUG]{ENTEG, Faculty of Science and Engineering, University of
Groningen, 9747 AG Groningen, The Netherlands}
\address[TUD]{Department of Mechanical Engineering, TU Dortmund University, Leonhard-Euler-Str. 2, Dortmund, 44227, Germany}

\begin{abstract}
 Search for the optimizer in computationally demanding model predictive control (MPC) setups can be facilitated by Cloud as a service provider in cyber-physical systems.
 This advantage introduces the risk that Cloud can obtain unauthorized access
to the privacy-sensitive parameters of the system and cost function. To solve this issue, i.e., preventing Cloud from accessing the parameters while benefiting from Cloud computation, random affine transformations provide an exact yet light weight in computation solution.
This research deals with analyzing privacy preserving properties of these transformations when they are adopted for MPC problems.
We consider two common strategies for outsourcing the optimization required in MPC problems, namely separate and dense forms, and establish that random affine transformations utilized in these forms are vulnerable to side-knowledge from Cloud. 
Specifically, we prove that the privacy guarantees of these methods and their extensions for separate form 
are undermined
when a mild side-knowledge about the problem in terms of structure of MPC cost function is available.
In addition, while we prove that outsourcing the MPC problem in the dense form inherently leads to some degree of privacy for the system and cost function parameters, we also establish that affine transformations applied to this form are nevertheless prone to be undermined by a Cloud with mild side-knowledge.
Numerical simulations confirm our results.
\end{abstract}
\end{frontmatter}
\endNoHyper

\section{Introduction}

Cloud computing allows accessing to computing resources such as networks, servers, storage, applications, and services that can be rapidly provisioned and released with minimal management effort or service provider interaction \cite{mell2011nist}. 
Researchers in control systems have reaped these benefits for a computationally demanding controller, namely model predictive control (MPC), across numerous domains where often interacting with geographically dispersed sensors and actuators are required.
For instance, Cloud-based MPC offers $53\%$ energy saving and $36\%$ improvement in thermal comfort for an office building \cite{Buildingdrgovna2020cloud}, consumption reduction of electrical energy by $71\%$ in geothermal fields \cite{GeoThermalStoffel2022cloud}, reducing the peak demand during cooling season by $24\%$ in public schools \cite{K12SchoolWoo2023first}, and flexibility of services in control of an industrial robot \cite{RobotVick2016model}.
Nonetheless, numerous security issues arise in Cloud based MPC due to the nature of Cloud computing, which among them is the unauthorized access of Could to private data such as system and controller parameters  \cite{xia2022brief}.
\mypar
To benefit from Cloud based MPC or in general Cloud based optimization \textit{and} to preserve the privacy of sensitive parameters against Cloud, numerous solutions have been proposed which can be categorized into \textit{three} main classes.
The \textit{first} class is based on the idea of using noise, where differential privacy (DP) is the main method in adjusting its amount \cite{dwork2014algorithmic}.
DP designs a mechanism for sufficiently perturbing data using noises with distributions such as Gaussian and Laplace before releasing the results of computation; for a tutorial in control systems, see \cite{cortes2016differential}.
While DP was used for private filtering in \cite{le2013differentially}, it was employed in \cite{hale2017cloud} to design a Cloud-enabled deferentially private multi-agent optimization. The works \cite{han2016differentially} and \cite{nozari2016differentially} also study DP for distributed optimizations, where perturbing the signals and objective functions are the tools in achieving DP.
Further on, the authors in \cite{yazdani2022differentially} and \cite{degue2022cooperative} utilize Cloud for designing a linear quadratic controller, where the first study focuses on system properties and the second is concerned with designing a filter before applying the privacy mechanism. 
\mypar
The \textit{second} class of methods for achieving privacy in Cloud based optimizations relies on adopting tools from cryptography, which among them Homomorphic Encryption (HE) is dominant. 
HE refers to cryptosystems that allow addition and/or multiplication of encrypted data without the need for decryption. While an optimization problem often needs to be modified to meet the framework of cryptosystems, the ability of computation over encrypted data enables the control designer to let Cloud securely solve the problem; for a control centered tutorial we refer to \cite{darup2021encrypted}. 
While a partially HE was used in \cite{gonzalez2014state} for the design of a privacy preserving Kalman filter and later on for encrypted control in \cite{kogiso2015cyber}, it was used in \cite{shoukry2016privacy} and \cite{alexandru2020cloud} to address the problem of privacy in quadratic optimizations where the latter also draws on tools from secure multiparty computation. 
Along the same line, the authors in \cite{alexandru2020towards} and \cite{lu2018privacy} adopted fully HE and leveled HE for data driven MPC and a Cloud based distributed optimizations.
In another line of research, the private explicit MPC was studied in \cite{darup2017towards} using mainly  Paillier encryption system, where later on the method was computationally improved in \cite{schluter2020encrypted}, and further was refined in \cite{tjell2021secure} using a cryptographic tool known as garbled circuits.
\mypar
Both of the classes have their own advantages and disadvantages:  differentially private algorithms offer privacy guarantees against Cloud (and other adversaries) with arbitrary computational power and side-knowledge \cite[p.237]{dwork2014algorithmic}, but in general they lead to suboptimal solutions where privacy-performance trade-offs need to be considered. On the other hand, cryptographic based methods do not lead to suboptimality and hence performance degradation \cite[p.7]{katz}, nonetheless they can be computationally and communication wise prohibitive for real time applications.
\mypar
To overcome these drawbacks, a \textit{third} class of methods have been used which in essence utilize some random transformations as privacy mechanism to change a problem into an equivalent problem before sharing it with Cloud.
One of the first studies in this research area (in computer science literature) is \cite{atallah2002secure} where the authors design local preprocessing (disguising) of data for problems such as matrix multiplication, matrix inversion and convolution of signals before sending the involved data to Cloud.
Along the same line, the authors in \cite{vaidya2009privacy}, \cite{mangasarian2011privacy} and \cite{dreier2011practical}  have studied preserving privacy for linear programs and in \cite{salinas2016efficient} and \cite{zhou2015outsourcing} for quadratic programs for large scale problems where in general they use affine transformation and permutation techniques to securely outsource the original problem. We further refer to the survey \cite{shan2018practical} for other studies.
\mypar
In control systems, the idea of using transformations for preserving privacy in optimization problems started with \cite{weeraddana2013per} where a unified framework based on equivalent optimizations was developed by including the previous studies such as \cite{vaidya2009privacy}, \cite{mangasarian2011privacy} and \cite{dreier2011practical}.
Further on, the approach was adopted in \cite{sultangazin2018towards}, \cite{sultangazin2019symmetries} and \cite{sultangazin2020symmetries} and tailored for optimization problems with quadratic costs and linear constraints. More specifically, an algorithm in \cite{sultangazin2020symmetries} for Cloud and system interactions was designed using  isomorphisms and symmetries of control systems to securely solve an MPC problem.
Moreover, the amount of guaranteed privacy was quantified by considering the dimension of the uncertainty set.
In further development, \cite{zhang2021privacy} and \cite{naseri2022privacy} have adopted the algebraic transformation method in \cite{sultangazin2020symmetries} for nonlinear and set theoretic MPC, meanwhile \cite{hayati2022privacy} and \cite{hayati2022immersion} have further improved it by adding structured noise to the signals in the context of federated learning and anomaly detection systems.
\mypar
\textit{Contribution.} In this research, we  analyze  the \textit{third} class of methods for preserving privacy in an MPC problem (more generally, quadratic optimizations). The study is along  \cite{schluter2023cryptanalysis} and \cite{Binfet} where the first investigates random affine transformations from cryptography perspective, and the second examines the invariant properties of quadratic optimizations under random affine transformations. 
Here, we take a different approach, namely we examine privacy preserving properties of random affine transformations 
 when Cloud has mild side-knowledge about the problem. This side-knowledge is based on common characteristics that can be found in control systems.
 We demonstrate that how this mild side-knowledge by Cloud weakens some of the privacy guarantees that these methods in general promise for system  and  controller parameters. 
 \mypar
 To this end, we begin with analyzing random affine transformations when they are applied to the  separate form of MPC. We prove that by using the structure of the MPC cost function, Cloud obtains unauthorized access to private information about the dynamic of the system under control and cost function parameters.
  Furthermore, we show that adding noise to the transformation, increasing its dimension, changing it to a nonlinear transformation or making it time-varying will not prevent Cloud from inferring privacy-sensitive parameters.
   In addition, we demonstrate that structured noise in high dimension 
   random affine transformations when it is adopted to preserve privacy for signals in linear systems is also vulnerable to side-knowledge.
  \mypar
In other direction, we address another strategy for securely outsourcing the solving of MPC problems
  and that is the case when the linear system is densely combined with the cost function and constraints and then random affine transformations are adopted. 
    We  prove that  combing a linear system with the quadratic cost in the MPC problem \textit{inherently} leads to a better privacy preserving situation for the cost function parameters (they are uncertain within a set). Nonetheless, we show that Cloud is still able to infer privacy-sensitive values about the system and cost function parameters (up to the inherent uncertainty set) by drawing on side-knowledge such as stability of the system.
   \mypar
\textit{Organization:} The rest of the paper is organized as follows: In Section \ref{sec:Pro_for} we present the problem formulation along the main assumption in the study, in Section \ref{sec:Pr_analysis_sep} we address the privacy preserving guarantees for affine transformations and its numerous extension when the system dynamics are separate from cost function.
 In Section \ref{sec:Pr_analysis_cmp_per_kno} we study the inherent privacy
 guarantees for the dense form of MPC, and in Section \ref{sec:affine_tra_com} we analyze the effects of side-knowledge on privacy guarantees of affine transformations. 
Numerical simulations and concluding remarks are presented in Section \ref{sec:num_sim} and \ref{sec:con}.
\mypar
\textit{Notation:} The set of positive, nonnegative integers and real numbers are denoted by $\N$, $\N_{0}$ and  $\R$, respectively. We denote the identity matrix of size $n$ by $I_n$, the zero matrix of size $n \times m$ by $0_{n \times m}$, the vector of size $n$ with $1$ in all its entries by $\bo_{n}$ and we drop the index whenever the dimension is clear from the context. By $A \succ0$($\succeq 0$), we mean $A$ is a positive (semi-) definite matrix. With $\otimes$, we denote the Kronecker product. For a matrix $A \in \R^{m \times n}$, $A^{\ell}$ is a left inverse ($A^{\ell}A = I_n$) when it exists, and $A^\dagger$ denotes its Moore-Penrose inverse. Given the elements $\{a_i\}_{i=1}^{n}$(scalars, vectors or matrices), we denote $\bbm a_1^\top, \ldots, a_n^\top \ebm^\top$ by $\col(a_1, \ldots, a_n)$ and the block matrix with $\{a_i\}_{i=1}^{n}$ on its diagonal by $\bdiag(a_1, \ldots, a_n)$. With bold symbols, we denote the secret in transformations.

\section{Problem formulation}\label{sec:Pro_for}
Consider a discrete-time LTI system 
\be\label{eq:lin_sys}
\begin{aligned}
	x_{k+1} &= Ax_k + Bu_k \\
	y_k&=Cx_k,
\end{aligned}
\ee
where $x_k \in \R^{n}$ is the state, $u_k \in \R^{m}$ the input,  $y_k \in \R^{p}$ the output at time $k \in \mathbb{N}_0$ and the matrices $A$, $B$, $C$ have consistent dimensions.
 In this study, we assume that the system \eqref{eq:lin_sys} is controllable and observable.
For system \eqref{eq:lin_sys}, we define the quadratic cost function over a finite horizon of $N$ steps as
\be\label{eq:LQR_cost}
J(\bx,\bu)\coloneqq x_N^{\top}Px_N + \sum_{k=0}^{N-1} \bbm x_k^{\top} & u_k^{\top} \ebm M  \bbm x_k \\u_k \ebm,
\ee 
where $M = \bdiag(Q, R)$ with $Q\succeq 0$, $R \succ 0$ and $P \succeq 0$ determine the cost function matrices and 
$\bx \coloneqq \col(x_0, x_1, \ldots, x_{N-1})$  and
$\bu \coloneqq \col(u_0, u_1, \ldots, u_{N-1})$.
In an MPC problem with bounds on inputs and outputs, we need to solve the following quadratic optimization problem \cite{borrelli2017predictive}:
\bse\label{eq:classic_mpc_cons}
\begin{align}
\min_{\bx, \bu} \quad & J(\bx,\bu)\\
\quad \text{s.t.} \quad &\eqref{eq:lin_sys}\\
&	u_{\text{min}} \leq u_k \leq u_{\text{max}}, & \quad  &k = 0, \ldots, N-1 \label{eq:classic_u_cons} \\
&	y_{\text{min}} \leq y_k \leq y_{\text{max}}, & \quad  &k = 1, \ldots, N, \label{eq:classic_y_cons}
\end{align}
\ese
where  $x_0 = x(0)$, $u_{\text{min}}, u_{\text{max}} \in \R^m$ and $y_{\text{min}}, y_{\text{max}} \in \R^p$ specify the constraints on $u_k$ and $y_k$, and 
$\bx$ and $\bu$ are the optimization variables.
After computing the optimizer $\bu^*$, MPC typically builds on applying the first block, i.e. $\bu^*_0$, to the system and repeating the optimization at the next sampling instance\cite[p. 89]{rawlings2017model}.
\mypar
The plant owner or simply Plant can outsource computation of the optimizer pair  $(\bx^*,\bu^*)$ to a service provider (Cloud) by requiring Cloud to follow a set of instructions (an algorithm). How Cloud behaves in the algorithm is determined below \cite[p.37]{cramer2015secure}:
\begin{definition}[Cloud model]\label{def:cloud_model} 
	Cloud is an honest-but-curious party, meaning that it follows any instructions given by Plant and does not deviate from them, but it stores the data received from Plant to learn more than what the algorithm allows.$\qedwhite$ 
\end{definition}
Note that Definition \ref{def:cloud_model} also specifies the adversary model in the language of secure control \cite{teixeira2015secure}. Furthermore, while Cloud in this study is a passive adversary, it  can cooperate with other type of adversaries by sharing sensitive information about Plant. 
 \mypar
 In general, Plant and Cloud can interact with each other for solving \eqref{eq:classic_mpc_cons} in two distinct ways:
\begin{enumerate}[(I)]
	\item \label{case_1} Plant  \textit{separately}  transmits to Cloud the system dynamic \eqref{eq:lin_sys}, cost function \eqref{eq:LQR_cost}, and constraints \eqref{eq:classic_u_cons}-\eqref{eq:classic_y_cons}. 
	\item \label{case_2} Plant \textit{densely} combines the linear system \eqref{eq:lin_sys} with cost function \eqref{eq:LQR_cost} and constraints  \eqref{eq:classic_u_cons}-\eqref{eq:classic_y_cons}, and then transmits them to Cloud.
\end{enumerate}
For each case, Plant uses random affine transformations ({to be determined later}) to conceal privacy-sensitive parameters of the system dynamics and cost from Cloud before outsourcing  the computation of the optimizer $(\bx^*,\bu^*)$ of \eqref{eq:classic_mpc_cons}. 
The central problem of interest that we aim to address in this manuscript is as follows:
\begin{problem}\label{q:main_question}
Suppose that Plant adopts random affine transformations as its privacy preserving mechanism. For each case (I) and (II), {determine}
what Cloud infers about system dynamics and cost parameters.  $\qedwhite$ 
\end{problem}
To answer Problem \ref{q:main_question} we make the following standing assumption throughout the manuscript:
\begin{sassumption}\label{assum:standing}
	Cloud is aware of the structure of  the outsourced problem as well as the adopted privacy-preserving mechanism, except for the values of the random affine transformations. $\qedwhite$ 
\end{sassumption}
The assumption is consistent with the standard assumption in cryptography, known as Kerckhoffs’ principle or Shannon’s maxim, that 
an encryption scheme must be designed to be secure even if the adversary knows all the details of the scheme \textit{except for the key} (see, e.g. \cite[p.7]{katz}). 
The other reason that urges us to work under this assumption is the role that knowledge of the system model plays in performing an attack against CPSs by adversaries.
As it has been reported in the literature of secure control (e.g. \cite{teixeira2015secure}) covert, zero dynamics and false data injection attacks to name but a few, rely largely on the knowledge that an adversary has about the system under control. Therefore, it is preferable to work with a comprehensive a priori knowledge for  Cloud (adversary), as in the case of Standing Assumption \ref{assum:standing}.

\section{Privacy Analysis of transformations for MPC in separate form}\label{sec:Pr_analysis_sep}
In this section, we provide an answer to Problem \ref{q:main_question} for the first case, i.e., when Plant transmits to Cloud cost and linear system separately after applying a transformation.
\subsection{Privacy mechanism}
To preserve the privacy of system and cost parameters in \eqref{eq:classic_mpc_cons}, Plant can use the following transformation 
\begin{subequations}\label{eq:iso}
	\begin{align}
	\tilde{x}_k &= \boldsymbol{T} x_k \label{eq:T_state} \\
	\tilde{u}_k &= \bF x_k+\bG u_k   \label{eq:T_input}\\
	\tilde{y}_k &= \bS y_k, \label{eq:T_output}
	\end{align}
\end{subequations}
where the \textit{randomly} chosen matrices $\bT \in \R^{n\times n}$, $\bG \in \R^{m \times m}$, and $\bS \in \R^{p \times p}$ are invertible, and $\bF\in \R^{m \times n}$.
Under the transformation \eqref{eq:iso}, the system dynamic is mapped to
\be\label{eq:lin_transformed}
\begin{aligned}
	\tilde{x}_{k+1} &= \underbrace{\bT(A-B\bG^{-1}\bF)\bT^{-1}}_{\eqqcolon\tilde{A}}\tilde{x}_k + \underbrace{\bT B\bG^{-1}}_{\eqqcolon\tilde{B}}\tilde{u}_k\\
	\tilde{y}_k &= \underbrace{\bS C\bT^{-1}}_{\eqqcolon\tilde{C}}\tilde{x}_k.
\end{aligned}
\ee  
Similarly, the cost function \eqref{eq:LQR_cost} is modified to
\be\label{eq:mod_cost}
\tilde{J}(\bxt, \but) \coloneqq \tilde{x}_N^\top\tilde{P}\tilde{x}_N + \sum_{k=0}^{N-1} \bbm \tilde{x}_k^{\top} & \tilde{u}_k^{\top} \ebm\tilde{M} \bbm \tilde{x}_k \\ \tilde{u}_k \ebm,
\ee
where $\tilde{P}\coloneqq \bT^{-\top} P\bT^{-1} $ and $\tilde{M}\coloneqq\bL^\top M \bL$ with
\be\label{eq:L_form_tr}
\bL\coloneqq \bbm \bT^{-1} & 0 \\
- \bG^{-1}\bF\bT^{-1} & \bG^{-1}
\ebm.
\ee
Note that we have not included \eqref{eq:classic_u_cons}-\eqref{eq:classic_y_cons} since our privacy analysis in this section is independent of those constraints. 
\mypar
The transformation \eqref{eq:iso} has been formally proposed in \cite{sultangazin2020symmetries} and based on it an algorithm for solving \eqref{eq:classic_mpc_cons} has been developed.
	In the algorithm, Cloud receives from Plant the transformed system \eqref{eq:lin_transformed} and the cost function \eqref{eq:mod_cost} in the start of their interactions (the handshaking phase) and then upon receiving $\tilde{y}_k = \bS y_k$ and  \textit{estimating} $\tilde{x}_k$ Cloud solves the optimization \eqref{eq:mod_cost} to obtain $\tilde{u}_k$. Then, Cloud sends $\tilde{u}_k$ to Plant where it can retrieve $u_k$ using the equality in \eqref{eq:T_input}. 
 
 \begin{remark}\label{rem:merits_of_iso}
 The privacy-preserving mechanism based on \eqref{eq:iso} is correct namely the optimizer of the original problem \eqref{eq:classic_mpc_cons} can be retrieved exactly (unlike most noise-based methods), and it is light-weight in computations, meaning Plant does not need to undergo extra computations other than the linear transformation in \eqref{eq:iso}. As such, the required computation is significantly lower than  what is needed for solving problem \eqref{eq:classic_mpc_cons}, in contrast to most standard cryptography-based methods. 
	It should be mentioned that the transformation and the analysis for its privacy guarantees are the tailored versions of transformations in \cite{weeraddana2013per}, \cite{zhang2021privacy} and \cite{naseri2022privacy} for quadratic optimizations. This means that while our analysis is concerned with  the transformation in \cite{sultangazin2020symmetries} it can be applied to the results in the related studies. $\qedwhite$
\end{remark}
\subsection{Privacy analysis}
Before proceeding with the analysis, we explicitly state the assumption below, with regard to Cloud's knowledge about the cost function. 
\begin{assumption}\label{assum:cloud_knowledge}
	Cloud knows that the matrix $M$ in \eqref{eq:LQR_cost} has a block diagonal \textit{structure}. $\qedwhite$ 
\end{assumption}
We emphasize that, consistent with Standing Assumption \ref{assum:standing}, Cloud \textit{does not} know the ground-truth value of $Q$ and $R$ in $M$. Note, the block diagonal $M$ corresponds to the conventional LQR problem \cite[p.33]{lewis2012optimal}.
  Next, we state how Cloud can use the side-knowledge in Assumption \ref{assum:cloud_knowledge} to infer privacy-sensitive information about the system dynamics and cost function.
\begin{theorem}\label{thrm:LQ_sep_main}
Let Assumption \ref{assum:cloud_knowledge} hold, and suppose that Cloud knows the transformed matrices $(\tilde A, \tilde B, \tilde C)$ in \eqref{eq:lin_transformed} and the transformed cost function in \eqref{eq:mod_cost}. Then, Cloud infers $\hat{A} = \bT A\bT^{-1}$ and $\hat{Q} = \bT^{-\top} Q\bT^{-1}$   about the system and cost parameters.
\end{theorem}
\emph{Proof}. See  Appendix \ref{app:proofs}.$\qed$
\mypar 
It follows from Theorem \ref{thrm:LQ_sep_main} that  under the mild side-knowledge in Assumption \ref{assum:cloud_knowledge}, Cloud can circumvent this transformation and essentially identify the dynamics up to a similarity $(\bT)$ as well as re-scaling transformations ($\bS$ and $\bG$).
For instance, using $\hat{A} = \bT A \bT^{-1}$, it finds the eigenvalues of the  state matrix $A$, i.e., poles of the system \eqref{eq:lin_sys}, which in turn can be leveraged by adversaries for designing additional cyber-attacks. 
\mypar
Next, we present a few remarks to further elaborate on different aspects of \eqref{eq:iso}:
\begin{remark}[Treatment of side-knowledge in \cite{sultangazin2020symmetries}]
	It is worth recasting the result of Theorem \ref{thrm:LQ_sep_main} in the framework of \cite{sultangazin2020symmetries}, where different scenarios concerning the side-knowledge available to Cloud are treated. 
	The side-knowledge in Assumption \ref{assum:cloud_knowledge} translates into the design constraint that  
	 Plant should ensure that the modified cost ($\tilde{M}$) in \eqref{eq:mod_cost}
	conforms with Cloud's knowledge. This means that the blocks $\tilde{M}_{2,1}$ and $\tilde{M}_{1,2}$ should be equal to $0$ which are satisfied if and only if $\bF=0$. Hence, the set of isomorphism in the terminology of \cite{sultangazin2020symmetries} is reduced to $\psi = (\bT,\boldsymbol{0},\bG,\bS)$, which weakens the privacy guarantees in the notable case of the cost function \eqref{eq:LQR_cost}.$\qedwhite$ 
\end{remark}

\begin{remark}[General cost function]\label{rem:gen_cost}
	It must be noted that when the matrix $M$ in cost function has \textit{no specific structure} Cloud cannot perform the inference attack described in Theorem \ref{thrm:LQ_sep_main}. To see this, let the matrix $M$ in \eqref{eq:LQR_cost} be redefined as
	$$
	M = \bbm Q & N \\ N^{\top} & R \ebm,
	$$
	where $M \succeq 0$ and $R\succ 0$. The matrix $\tilde{M}$ in the transformed cost \eqref{eq:mod_cost} is then given by:
	\[
	\tilde{M} = \bbm \tilde{M}_{1,1} & \star \\ \bG ^{-\top}N^\top \bT^{-1} -(\bG^{-\top}R\bG^{-1})\bF\bT^{-1} & \bG^{-\top}R\bG^{-1} \ebm,
	\]
	with
	\begin{align*}
	\tilde{M}_{1,1} =  \bT^{-\top} &Q\bT^{-1} + (\bF \bT^{-1})^{\top}(\bG^{-\top}R\bG^{-1})(\bF \bT^{-1}) \\- &\bT^{-\top}N\bG^{-1}\bF \bT^{-1} -(\bF \bT^{-1})^{\top}\bG^{-\top}N^{\top}\bT^{-1} .
	\end{align*}	 
The additional term $\bG ^{-\top}N^\top \bT^{-1}$ in block $(2,1)$ hinders Cloud from applying the sequence of observations detailed in the proof of Theorem \ref{thrm:LQ_sep_main}. In particular, the matrix $\bF \bT^{-1}$ cannot be computed in the second step. $\qedwhite$
\end{remark}
\begin{remark}[Loss of observability]\label{rem:obs}
Note that if $(A,B,C)$ is controllable and observable, the transformed system $(\tilde{A}, \tilde{B}, \tilde{C})$ in \eqref{eq:lin_transformed} is controllable but not \textit{necessarily} observable. This is due to the fact that the artificially introduced state feedback \eqref{eq:T_input} preserves controllability, but not necessarily observability.
Bearing in mind the  discussion preceding Remark \ref{rem:merits_of_iso}, state estimation is a part of the protocol outsourced to  Cloud. Hence, loss of observability has a major consequence in the execution of the algorithm. 
This means that $\bF$ cannot be picked arbitrarily; but rather it should be selected such that the transformed pair $(\bT(A-B\bG^{-1}\bF)\bT^{-1}, \bS C\bT^{-1} )$ is observable.  
An example is provided next. 
$\qedwhite$ 
\end{remark}
\begin{example}
Consider a system given by the triple:
	$$
	C = \bbm1 & 0 \ebm, A = \bbm 0 & 1 \\ 1 & 0\ebm, B = \bbm 1 \\ 0\ebm,
	$$
	which is both observable and controllable.
	Consider a random transformation according to \eqref{eq:iso} with
	$\bF = [ 0 \quad f ]$, $\bG = 1$, $\bT = I_2$, $\bS =1$, and some scalar $f$.
The transformed system given by
	$$
	\tilde{C} = \bbm1 & 0 \ebm, \tilde{A} =\bbm  0 & 1-f \\ 1 & 0\ebm, \tilde{B} = \bbm 1 \\ 0\ebm,
	$$
	is controllable, but it is observable if only if $f\ne 1$.  $\qedwhite$ 
\end{example}
Next, we show that even by using an affine form of transformation \eqref{eq:iso} or increasing its dimension, Cloud is still able to infer sensitive information.
\subsection{Other forms of transformation \eqref{eq:iso}}
We address the privacy analysis when Plant utilizes some variants of  transformation \eqref{eq:iso} for preserving privacy of linear system and cost function parameters.
\subsubsection{Affine form of \eqref{eq:iso}}
Let Plant adopt
the following transformation
	\be\label{eq:affine_tr}
		\begin{aligned}
		\tilde{x}_k &= \boldsymbol{T} x_k + \bro  \\
		\tilde{u}_k &= \bF x_k+\bG u_k + \brt   \\
		\tilde{y}_k &= \bS y_k + \brth,
		\end{aligned}
	\ee
	where $\bro \in \R^{n}$, $\brt \in \R^{m}$, and $\brth \in \R^{p} $ are random vectors and the matrices $\bT$, $\bF$, $\bG$ and $\bS$ are as in \eqref{eq:iso}. Note that the difference with \eqref{eq:iso} arises from the vectors $\bro$, $\brt$, and $\brth$.
 Under the transformation \eqref{eq:affine_tr}, the linear system \eqref{eq:lin_sys} is modified to
	\be\label{eq:aff_tr_system}
	\begin{aligned}
		\tilde{x}_{k+1} &=\tilde{A}\tilde{x}_k + \tilde{B}\tilde{u}_k+ \bT A r_x + \bT B r_u + \bro \\
		\tilde{y}_k & =  \tilde{C}\tilde{x}_k + \bS C r_x + \brth,
	\end{aligned}
	\ee
	where $\tilde{A}$, $\tilde{B}$, and $\tilde{C}$ are given in \eqref{eq:lin_transformed} and $r_x\coloneqq- \bT^{-1}\bro$ and $r_u \coloneqq\bG^{-1}\bF\bT^{-1} \bro -  \bG^{-1}\brt$. Similarly, the cost function \eqref{eq:LQR_cost} is modified to
	\be\label{eq:aff_tr_cost}
	\begin{aligned}
		&\tilde{J}_{r}(\bxt, \but)\coloneqq \tilde{J}(\bxt, \but) + 2 \sum_{k=0}^{N-1} \bbm \tilde{x}_k^{\top} & \tilde{u}_k^{\top} \ebm \bL^\top M \bbm r_x \\ r_u \ebm  \\
		& + 2\tilde{x}_N^\top \bT^{-\top}P r_x 
		+ r_x^\top P r_x + N \bbm r_x^{\top} & r_u^{\top} \ebm M \bbm r_x \\ r_u \ebm, 
	\end{aligned}
	\ee
	where $\tilde{J}(\bxt, \but)$ and $\bL$ are given in \eqref{eq:mod_cost} and  \eqref{eq:L_form_tr}. We have the following result: 
	\begin{corollary}\label{cor:affine}
		 Let Assumption \ref{assum:cloud_knowledge} hold, and suppose that Cloud knows the transformed system \eqref{eq:aff_tr_system} and cost function \eqref{eq:aff_tr_cost}.
	 Cloud infers $\hat{A} = \bT A\bT^{-1}$ and $\hat{Q} = \bT^{-\top} Q\bT^{-1}$   about the system and the cost parameters.
	\end{corollary}
\emph{Proof}. See  Appendix \ref{app:proofs}.$\qed$

\subsubsection{{Increasing the dimension  of \eqref{eq:iso}}}Let Plant adopt for the optimization \eqref{eq:classic_mpc_cons} the following higher-dimensional transformation
\begin{subequations}\label{eq:T_high}
	\begin{align}
	\tilde{x}_k &= \cbT x_k  \label{eq:high_T_state} \\
	\tilde{u}_k &= \cbF x_k+\cbG u_k  \\
	\tilde{y}_k &= \cbS y_k \label{eq:high_T_output}, 
	\end{align}
\end{subequations}
where $\cbT \in \R^{\bar n \times n}$, $\cbG \in \R^{\bar m \times m}$, $\cbS \in \R^{\bar p \times p}$ are randomly chosen full column rank matrices with $\bar n > n$, $\bar m > m$, and $\bar p > p$, and $\cbF \in \R^{\bar{m} \times n}$. Under this transformation, the system \eqref{eq:iso} turns into
\be\label{eq:sys_T_high}
\begin{aligned}
	\tilde{x}_{k+1} &= \underbrace{\cbT (A - B\cbG^{\ell}\cbF)\cbT^{\ell}}_{\eqqcolon\tilde{A}}\tilde{x}_k + \underbrace{\cbT B\cbG^{\ell}}_{\eqqcolon\tilde{B}}\tilde{u}_k\\
	\tilde{y}_k &= \underbrace{\cbS C\cbT^{\ell}}_{\eqqcolon\tilde{C}}\tilde{x}_k,
\end{aligned}
\ee
where  $\cbT^{\ell}$, $\cbG^{\ell}$ and $\cbS^{\ell}$ are \textit{any} left inverse of their corresponding matrices. The cost function \eqref{eq:LQR_cost} is transformed into
\be\label{eq:cost_T_high}
\tilde{J}_{h}(\tilde{\bx}, \tilde{\bu}) \coloneqq \tilde{x}_N^\top (\cbT^{\ell})^{\top} P\cbT^{\ell}\tilde{x}_N + \sum_{k=0}^{N-1} \bbm \tilde{x}_k^{\top} & \tilde{u}_k^{\top} \ebm\tilde{M} \bbm \tilde{x}_k \\ \tilde{u}_k \ebm,
\ee
where $\tilde{M}$ is given by
\be\label{eq:M_tilde_high}
\begin{aligned}
	\tilde{M} = \bbm \tilde{M}_{1,1} & \star \\ -((\cbG^{\ell})^{\top}R\cbG^{\ell})(\cbF\cbT^{\ell}) & (\cbG^{\ell})^{\top}R\cbG^{\ell} \ebm,
\end{aligned}
\ee
with 
$
\tilde{M}_{1,1} = (\cbT^{\ell})^{\top} Q\cbT^{\ell} + (\cbF\cbT^{\ell})^{\top}((\cbG^{\ell})^{\top}R\cbG^{\ell})(\cbF\cbT^{\ell})
$, and $(\star)$ denotes the transpose of block $\tilde{M}_{2,1}$.
\begin{corollary}\label{cor:high_dim_sep}
		 Let Assumption \ref{assum:cloud_knowledge} hold, and suppose that Cloud knows the transformed system \eqref{eq:sys_T_high} and cost function \eqref{eq:cost_T_high}.
	Cloud infers $\hat{A} = \cbT A\cbT^{\ell}$ and $\hat{Q} = (\cbT^{\ell})^{\top} Q\cbT^{\ell}$ about the system and cost parameters.
	Furthermore, Cloud infers $\mspec_{nz} (A)$, where $\mspec_{nz} (A)$ denotes the multiset of nonzero eigenvalues of $A$.
\end{corollary}
\emph{Proof}. See  Appendix \ref{app:proofs}.$\qed$
\mypar
We present two other forms of \eqref{eq:iso} in two separate remarks. 
\begin{remark}[Time-varying high dimensional affine transformations]\label{rem:time_var_tra}
	It should be noted that Plant can adopt for each instance of optimization \eqref{eq:classic_mpc_cons} a time dependent transformation given by quadruple $(\bT_k, \bG_k, \bS_k, \bF_k)$ in \eqref{eq:iso} justified by the recommendation in privacy literature to use random transformation only once \cite{shan2018practical}.  Nonetheless, it is clear that Cloud can still follow the steps in Theorem \ref{thrm:LQ_sep_main} mainly because it does not rely on accumulated information from previous interactions with Plant and hence it is able to infer $\hat{A} = \bT_k A\bT_k^{-1}$ and $\hat{Q} = \bT_k^{-\top} Q\bT_k^{-1}$ and $\hat{R}=\bG_k^{-\top}R\bG_k^{-1}$. We also point out that using time-varying transformation along with added randomly chosen vectors and increasing its dimension as  in \eqref{eq:affine_tr} and \eqref{eq:T_high} does not prevent Cloud from applying the arguments in Corollaries \ref{cor:affine} and \ref{cor:high_dim_sep} to infer privacy-sensitive quantities of the system. $\qedwhite$
\end{remark}
\begin{remark}[Polynomial form of \eqref{eq:iso}]
	Consider the following nonlinear transformation
	\be \label{eq:nonlinear_input}
	\tilde{u}_k = \bbf(x_k) + \bG u_k,  
	\ee
	where $\bbf(\cdot): \R^{n} \to \R^{m}$ is a polynomial function of degree $d \in \N$. The map $\bbf(\cdot)$ can be rewritten as
	$$
	\bbf(x) = \brt + \bF x + \bF_1 Z(x), 
	$$
 for some constant vector $\brt \in \R^{m}$ and matrices 
$\bF \in \R^{m \times n}$, $\bF_1 \in \R^{m \times z}$, with $Z(x): \R^{n} \to \R^{z}$ including all the $z$ monomials with degree greater than $1$. Let Plant use the transformation \eqref{eq:iso} with \eqref{eq:T_input} replaced by \eqref{eq:nonlinear_input}, where the vectors/matrices $\brt$, $\bF_1$, and $\bF$ are randomly chosen.
Under this transformation, the linear system \eqref{eq:lin_sys} is  modified to
	\be\label{eq:non_tr_system}
	\begin{aligned}
		\tilde{x}_{k+1} & = \tilde{A}\tilde{x}_k + \tilde{B}\tilde{u}_k + \bT B \phi(\tilde{x}_k) +\bT B r_u \\
		\tilde{y}_k & =  \tilde{C}\tilde{x}_k,
	\end{aligned}
	\ee
	where $\tilde{A}$, $\tilde{B}$, and $\tilde{C}$ are given in \eqref{eq:lin_transformed} and $\phi(\tilde{x}_k)\coloneqq -\bG^{-1}\bF_1 Z(\bT^{-1} \tilde{x}_k)$ and $r_u\coloneqq  - \bG^{-1}\brt$.
	Analogously, the transformed cost is:
	\begin{align*}
			\tilde{J}_{nl}(\bxt, \but)&\coloneqq \tilde{J}(\bxt, \but) \\
	&  + 2 \sum_{k=0}^{N-1} \bbm \tilde{x}_k^{\top} & \tilde{u}_k^{\top} \ebm \bbm  -(R\bG^{-1}\bF\bT^{-1})^\top \\ \bG^{-\top}R \ebm \phi(\tilde{x}_k)\\ 
			& + 2 \sum_{k=0}^{N-1} \bbm \tilde{x}_k^{\top} & \tilde{u}_k^{\top} \ebm \bbm  -(R\bG^{-1}\bF\bT^{-1})^\top \\ \bG^{-\top}R \ebm r_u
	\end{align*}
	\be\label{eq:non_tr_cost}
	\begin{aligned}
		&  + \sum_{k=0}^{N-1} \phi(\tilde{x}_k)^\top R \phi(\tilde{x}_k) + {2r_u^\top R \phi(\tilde{x}_k)} + r_u^\top R r_u,
	\end{aligned}
	\ee
	where $\tilde{J}(\bxt, \but)$ is given by \eqref{eq:mod_cost}. 
\mypar
 Bearing in mind that $\phi(\cdot)$ contains only high degree monomials,  the values of the matrices $\tilde{A}$, $\tilde{B}$ can be inferred from \eqref{eq:non_tr_system}.
 Furthermore, due to the structure of \eqref{eq:non_tr_cost}, 
 Cloud can isolate the terms $-\tilde{x}_k^\top(\bG^{-\top}R\bG^{-1})(\bF\bT^{-1})\tilde{u}_k$ and $\tilde{u}_k^\top\bG^{-\top}R\bG^{-1}\tilde{u}_k$, embedded in $\tilde{J}(\bxt, \but)$, from the rest of the terms in \eqref{eq:non_tr_cost}, and subsequently infer the values of $\hat{R}=\bG^{-\top}R\bG^{-1}$ and $\bF\bT^{-1}$. The latter, using the expression of $\tilde A$, reveals the matrix $\hat{A} = \bT A\bT^{-1}$ to  Cloud.  
$\qedwhite$
\end{remark}

The previously discussed results concerned with system level privacy along with privacy of the cost parameters. In certain applications, privacy of input and/or output signals should be protected.
While those scenarios do not directly deal with Problem \ref{q:main_question}, the side-knowledge analysis pursued here can shed some lights on the information leaked to Cloud. An example of such scenarios is provided in the remark below. 

\begin{remark}[Structured noise and signal privacy]
 Consider a transformation of $(x_k, y_k, u_k)$ given by \eqref{eq:high_T_state}, \eqref{eq:high_T_output} and 
	\be \label{eq:high_C_input}
	\tilde{u}_k = \cbG u_k + \cbb_k, \quad \text{with} \quad \cbb_k \in \ker \cbG^{\ell},
	\ee
	where $\cbG \in \R^{\bar m \times m}$ has full column rank with  $\bar m > m$. The vector $\cbb_k$ is selected  randomly from $\ker\cbG^{\ell}$ by Plant.
	The system dynamic \eqref{eq:lin_sys} under transformation \eqref{eq:high_T_state}, \eqref{eq:high_T_output} and \eqref{eq:high_C_input} is mapped to
	\be\label{eq:lin_trans_high}
	\begin{aligned}
		\tilde{x}_{k+1} &= \underbrace{\cbT A\cbT^{\ell}}_{\eqqcolon{A_h}}\tilde{x}_k + \underbrace{\cbT B\cbG^{\ell}}_{\eqqcolon{B_h}}\tilde{u}_k\\
		\tilde{y}_k &= \underbrace{\cbS C\cbT^{\ell}}_{\eqqcolon{C_h}}\tilde{x}_k.
	\end{aligned}
	\ee
	The immersed dynamic \eqref{eq:lin_trans_high} is transmitted to Cloud for computing the output $\tilde{y}$. For doing so, Cloud first receives the value of matrices $A_h$, $B_h$, $C_h$ and $\tilde{x}_{0} = \cbT x_0$ and later on, it receives the signal $\tilde{u}_k$ which is a randomly encoded version of $u_k$ and performs the computation of $\tilde{y}_k$.
The random transformation including \eqref{eq:high_T_state}, \eqref{eq:high_T_output} and \eqref{eq:high_C_input} has been proposed in \cite{hayati2022privacy}  and \cite{ hayati2022immersion} in the context of privacy preserving federated learning and detector systems.
	Note, under this transformation, the signal $\tilde{u}_k$ constructed by the random mapping \eqref{eq:high_C_input} ($\cbb_k$ is a random value for each $k$) can differ significantly from plaint-text $u_k$ in \eqref{eq:T_input}.
	This feature makes the transformed system \eqref{eq:lin_trans_high} a suitable candidate for  applications such as securely implementing a linear dynamical controller over Cloud similar to \cite{kim2022dynamic}.
 However, the effects of the noise $\cbb_k$ can be removed by having a mild assumption that
	the matrix $B$ in the system \eqref{eq:lin_trans_high} is full column rank
	and Cloud \textit{knows} this.
 \mypar
Under this assumption, it follows that  $ \ker \cbG^{\ell}= \ker B_h$ with $B_h = \cbT B\cbG^{\ell}$ in \eqref{eq:lin_trans_high}.
Hence, Cloud selects an arbitrary basis for $\ker \cbG^{\ell}$ as
	$
	\cbG^{\ell}_{kn}  \coloneqq  \bbm v_1 & v_2 & \cdots & v_{r}\ebm \in \R^{\bar{m} \times r}$, and consistently a basis for row space of $\cbG^{\ell}$ as $\cbG^{\ell}_{rs} \in \R^{(\bar{m}-r) \times \bar{m}}$.
	Note $\cbG^{\ell}$ must be full row rank since from its definition it holds that $\cbG^{\ell}\cbG = I_m$. 
	 Therefore, Cloud finds $\bar{m}-r \eqqcolon m$ as the hidden underlying number of rows of $\cbG^{\ell}$  and hence number of columns of $\cbG$.
	 Thus, Cloud can write $\cbG^{\ell}_{rs} = \cbQ \cbG^{\ell}$ for some unknown invertible $\cbQ\in \R^{m \times m}$ \cite[p.13]{horn2012matrix}. Multiplying both sides of \eqref{eq:high_C_input} from the left by $\cbG^{\ell}_{rs}$ yields
	\be\label{eq:Quk}
	\cbG^{\ell}_{rs}\tilde{u}_k = \cbQ u_k,
	\ee
	where Cloud knows its left-side. From \eqref{eq:Quk} we can see that Cloud faces $\cbQ u_k$ as the ultimate uncertainty with regard to $u_k$ when it receives ``noisy'' $\tilde{u}_k$. This means that $u_k$ becomes known to Cloud up to a constant matrix.$\qedwhite$
\end{remark}
In the rest of the study we provide an answer to Problem \ref{q:main_question} for the second case, namely when Plant \textit{densely} combines the linear system \eqref{eq:lin_sys} with the cost function \eqref{eq:LQR_cost} and constraints \eqref{eq:classic_u_cons}-\eqref{eq:classic_y_cons}, and then transmits them to Cloud.

\section{ Analysis of inherent privacy for MPC in dense form}\label{sec:Pr_analysis_cmp_per_kno}
First, we present the dense optimization form of \eqref{eq:classic_mpc_cons}. 

\subsection{Dense form of MPC}
We can rewrite the optimization problem \eqref{eq:classic_mpc_cons} 
  by expressing all future states $x_1, x_2, \ldots, x_N$ as the function of future inputs $u_0, u_1, \ldots, u_{N-1}$ and initial condition $x_0$. This gives rise to
  \cite{borrelli2017predictive}
\bse\label{eq:LQ_merged}
\begin{align}
	\min_{z} \quad &\frac{1}{2}z^{\top} H z + x_0^{\top}Fz
+ { x_0^{\top} Yx_0} \label{eq:cost_LQ_merged} \\
\quad \text{s.t.} \quad & Gz \leq W + Ox_0, \label{eq:constraint_LQ_merged}
\end{align}
\ese
where $
z \coloneqq \col(u_0, u_1, \ldots, u_{N-1}) \in \R^{Nm}
$ is the optimization variable, and $H\coloneqq 2{\calR} + 2{\calS}^{\top}{\calQ}{\calS}$, $F\coloneqq 2{\calT}^{\top}{\calQ}{\calS}$, and $Y\coloneqq Q+{\calT}^\top{\calQ}{\calT}$ are the cost matrices in \eqref{eq:cost_LQ_merged} with 
\be\label{eq:cost_all_block_matrices}
\begin{aligned}
	& {\calR} \coloneqq I_N \otimes R, \,\,
	{\calQ} \coloneqq \bdiag({Q}, \cdots, {Q} ,{P}) \in \R^{N(n \times n)}, \\
	&{\calS} \coloneqq \bbm B & 0 & \cdots & 0 \\
	AB & B & \cdots & 0 \\
	\vdots & \vdots & \ddots & \vdots \\
	A^{N-1}B & A^{N-2}B & \cdots & B \ebm, \quad
	{\calT} \coloneqq \bbm A \\ A^{2} \\ \vdots \\ A^{N} \ebm.
\end{aligned}
\ee
The expanded form of $H$, $F$ and $Y$ are also given in Appendix \ref{app:HFY}. The matrices in the constraint \eqref{eq:constraint_LQ_merged} are
\be\label{eq:constraint_all_block_matrices}
\begin{aligned}
	\calG &\coloneqq (I_{N}\otimes C)\calS, \,\, G \coloneqq \col(I_{Nm}, -I_{Nm}, {\calG} , -{\calG}), \\
	W &\coloneqq \bbm \begingroup\color{white}-\endgroup\bo_{N} \otimes u_{\text{max}} \\ -\bo_{N} \otimes u_{\text{min}} \\\begingroup\color{white}-\endgroup\bo_{N} \otimes y_{\text{max}} \\ -\bo_{N} \otimes y_{\text{min}}\ebm, \,\,  {\calO} \coloneqq \bbm CA \\ CA^{2} \\ \vdots \\ CA^{N} \ebm, \,\, 
	 O \coloneqq \bbm 0_{Nm \times n} \\ 0_{Nm \times n} \\ -{\calO} \\ \begingroup\color{white}-\endgroup{\calO} \ebm.	
\end{aligned}
\ee
We are interested in the knowledge that Cloud obtains about the system and cost matrices by using the received parameters from Plant.
 For doing so, we make the following assumption for the rest of the manuscript:
\begin{assumption} \label{Assum: A_inv} The state matrix
	$A \in \R^{n \times n}$ in \eqref{eq:lin_sys} is non-singular.$\qedwhite$
\end{assumption}
This assumption always holds in the case where matrix $A$ is obtained as a result of discretizing a linear continuous time system
\cite[p.80]{lewis2012optimal}.
\mypar
To determine the benefits of random affine transformations for the dense form of MPC \eqref{eq:LQ_merged}, we first examine (any) inherent privacy guarantees of the scheme, i.e., those without performing any privacy-preserving mechanism other than the local computations that led to the dense form \eqref{eq:LQ_merged}.
To this end, we distinguish between several cases based on the knowledge of Cloud on the parameters in \eqref{eq:LQ_merged}.

 \subsection{Privacy analysis when Cloud knows  $H$, $F$, $G$, $W$ and $O$ in  \eqref{eq:LQ_merged}}
In the first case, we address the privacy analysis when Plant transmits $H$, $F$, $G$, $W$ and $O$ to Cloud for solving \eqref{eq:LQ_merged}. 
The result of this case is summarized in the following theorem:
\begin{theorem}\label{thrm:GOHF_known}
	Suppose that Cloud knows the matrices $H$, $F$, $G$, $W$ and $O$ in  \eqref{eq:LQ_merged}. 
	If $N > n$ then Cloud recovers the system matrices $A$, $B$, $C$ in \eqref{eq:lin_sys} and the matrix $R$ in \eqref{eq:LQR_cost}.
	Furthermore, from the obtained set of matrices, Cloud infers the  cost matrices $Q$ and $P$ 
 if and only if the following set is a singleton:
	\be\label{eq:J_theorem}
	\begin{aligned}
		\calJ = \big\{X \in \R^{n \times n}|\, \, \, \hat{Q} + X - A^\top X A & \succeq 0, \\
		\hat{P} + X &\succeq 0, \\
		XB &= 0 \big\},
	\end{aligned}
 \ee
 where $(\hat{Q}, \hat{P})$ are any positive semi-definite matrices consistent with the structure of the matrices $H$ and $F$ in \eqref{eq:LQ_merged}.\footnote{More explicitly, the pair $(\hat{Q}, \hat{P})$ can be picked from the set given in \eqref{eq:set_Q_P_original}.}
\end{theorem}
\emph{Proof}. See  Appendix \ref{app:proofs}.$\qed$

\begin{remark}[privacy of $P \succeq 0$ and $Q \succeq 0$]\label{rem:P_Q_unique}
	It should be noted that in control systems the matrix $B$ is generally not invertible, namely $B \in \R^{n \times m}$, $\rank B = m$, and $m < n$.
	In that case, if the cost matrices $P$ and $Q$ %
 are positive definite, it is straightforward to see that the set $\calJ$ in \eqref{eq:J_theorem} is not a singleton. To see this, note that for an arbitrary $X$ with $XB=0$,  we have $P + \epsilon X \succeq 0 $ and $Q + \epsilon (X - A^\top X A) \succeq 0$ for sufficiently small $\epsilon$. 
From the perspective of privacy, this means that the dense form of MPC given in \eqref{eq:LQ_merged}  \textit{inherently} enjoys a certain degree of privacy for the cost matrices $P$ and $Q$ \textit{even if} the matrices $A$, $B$, and $R$ are publicly known. An analogous argument also applies to the case $P \succeq 0$, $Q \succ 0 $. $\qedwhite$ 
\end{remark}

\subsection{Privacy analysis when Cloud knows $H$, $x_0^\top F$, $G$ and $W+Ox_0$ in  \eqref{eq:LQ_merged}}
We address the case when Plant merges $x_0$ with $F$ and $O$ matrices in \eqref{eq:LQ_merged} and transmits $x_0^\top F$ and $W+Ox_0$ along $H$ and $G$ to Cloud.
To do so, first, we analyze the values that can be obtained from the matrix $G$.
Recall that the Markov parameters of the system are defined as \cite[p. 72]{verhaegen2007filtering}
\be\label{eq:markov}
h(k)  \coloneqq  CA^{k-1}B, \quad k=1,2, \ldots\,\, . 
\ee
It is clear that Cloud has access to Markov parameters through $G$ from which it is \textit{well-known} that the system parameters can be recovered under certain conditions.
\begin{lemma}[Identifying $(A, B, C)$ from $G$]\label{lem:G_known}
	Suppose that Cloud knows the matrix $G$ in \eqref{eq:LQ_merged}. 
	If $N \geq 2n+1$, then Cloud infers $A_t = T^{-1}AT$, $B_t = T^{-1}B$ and $C_t = CT$ about
	the system matrices with known $A_t$, $B_t$, $C_t$ and an unknown non-singular matrix $T \in \R^{n \times n}$.
\end{lemma}
\emph{Proof}. See  Appendix \ref{app:proofs}.$\qed$ 
\mypar
By Lemma \ref{lem:G_known}, the system matrices are revealed to Cloud up to a similarity transformation. We will show next that, under an additional assumption, some information about the cost matrices are also revealed. 

\begin{assumption}\label{Assum: multiple_MPC}
	The MPC problem \eqref{eq:LQ_merged} is solved for $I \in \mathbb{N}$ instances $\calI = \{1, 2, 3, \ldots, I\}$ with the same matrices $H$, $F$, $G$, $W$, and $O$ and (different) initial states $x_{0|1}$, $x_{0|2}$, $\ldots$, $x_{0|I}$. $\qedwhite$ 
\end{assumption}
The assumption is consistent with an MPC setup where Plant applies the first step of $z^*$ (the optimizer in \eqref{eq:LQ_merged}) 
to the system \eqref{eq:lin_sys} and then asks Cloud for a new optimizer based on the current state of the system.
\mypar
To state the next result, we collect the  initial states in the following  matrix
\be\label{eq:all_init}
X_0 \coloneqq\bbm x_{0|1} & x_{0|2} & \ldots & x_{0|I} \ebm \in \R^{n \times I}.
\ee
\begin{theorem}\label{thrm:GWOx0HFx0}
	Let Assumption \ref{Assum: multiple_MPC} hold, and 	suppose that Cloud knows the matrices $H$, $x_0^\top F$, $G$ and $W+Ox_0$ in \eqref{eq:LQ_merged}.
If $N \geq 2n+1$, $\rank X_0 = n$, and 
	$\bo_I \notin \im X_0^\top$,
 then Cloud recovers the system matrices $A = TA_tT^{-1}$, $B = TB_t$ and $C = C_tT^{-1}$ 
	with known $A_t$, $B_t$, $C_t$ and an unknown non-singular matrix $T \in \R^{n \times n}$, and identifies the  cost matrix $R$. 
Furthermore, from the obtained set of matrices, Cloud infers 
$Q = T^{-\top}Q_tT^{-1}$ and $P=T^{-\top}P_tT^{-1}$
if and only if the following  set is a singleton:
	\be\label{eq:J_theorem_up_to_sim}
	\begin{aligned}
		\calJ_t = \big\{X \in \R^{n \times n}|\, \, \, Q_t + X - A_t^\top X A_t & \succeq 0, \\
		P_t + X &\succeq 0, \\
		XB_t &= 0 \big\},
	\end{aligned}
	\ee
where $(Q_t, P_t)$ are any positive semi-definite matrices consistent with the structure of the matrices $H$ and $F$ in \eqref{eq:LQ_merged}.\footnote{More explicitly, the pair $(Q_t, P_t)$ can be picked from the consistency set given by \eqref{eq:set_Qt_Pt_original} in Appendix \ref{app:proofs}}
\end{theorem}
\emph{Proof}.
See  Appendix \ref{app:proofs}.	 $\qed$ 
\mypar
It follows from Theorem \ref{thrm:GWOx0HFx0} that Cloud can recover (depending on the set \eqref{eq:J_theorem_up_to_sim}) the system and cost matrices up to an unknown invertible matrix $T$ as
\begin{align*}
& \big(A, B, C\big) = \big(TA_tT^{-1}, TB_t, C_tT^{-1}\big) \\ 
&\big(P, Q, R \big) = \big(T^{-\top}P_tT^{-1}, T^{-\top}Q_tT^{-1}, R \big).
\end{align*}
Interestingly, this is the same information that Cloud obtains when Plant adopts the transformation given in \eqref{eq:iso} with $G = I_m$ and $S =I_p$. This means that  transmitting the matrices $H$, $x_0^\top F$, $G$ and $W+Ox_0$ to Cloud leads to the same privacy guarantees as in Theorem \ref{thrm:LQ_sep_main}.
\begin{remark}
    It should be noted that the conditions $\rank X_0 = n$ and 
$\bo_I \notin \im X_0^\top$
with $X_0$ in \eqref{eq:all_init} in Theorem \ref{thrm:GWOx0HFx0} are satisfied only if $I \geq n+1$. This imposes a lower bound on the number of instances in Assumption \ref{Assum: multiple_MPC}.
\end{remark}

\subsection{Privacy analysis when Cloud knows $H$ and $F$ in  \eqref{eq:LQ_merged}} \label{subsec:H_F_case}
We analyze the situation when Cloud has access to the matrices $H$ and $F$ (or $x_0^\top F$) in \eqref{eq:LQ_merged}. This case most notably arises when Plant reorders the constraints in \eqref{eq:LQ_merged} (as a computationally light privacy mechanism) since as we have shown in Lemma \ref{lem:G_known} 
the constraints  reveal the system dynamics. Another notable example is the case where the constraints are not present in \eqref{eq:LQ_merged}, i.e., there are no a priori bounds on input and output in \eqref{eq:classic_mpc_cons}. 
For treating this case, we make the following assumption:
\begin{assumption}\label{Assum:sys_stable} The following conditions hold:
	\begin{enumerate}[(i).]
		\item The matrix $A$ in \eqref{eq:lin_sys} is Schur stable
		\item The horizon parameter $N \to \infty$ in \eqref{eq:LQR_cost}$\qedwhite$ 
	\end{enumerate}
\end{assumption}
As for the second part of the assumption, consideration of control properties in MPC generally favors large horizon $N$ \cite[p. 374]{rawlings2017model}.
We further elaborate on this assumption and the roles that non-minimum phase zeros play for the MPC problem in Section \ref{sec:num_sim}.
\mypar 
Under Assumption \ref{Assum:sys_stable}, the matrix $Y = { Q+{\calT}^\top{\calQ}{\calT}}$ in cost function \eqref{eq:cost_LQ_merged}, i.e.,
$
Y = \sum_{i=0}^{N-1}{(A^{\top})^{i}QA^{i}} + (A^{\top})^{N}PA^{N},
$
converges to a matrix $\bY \in \R^{n \times n}$  \cite[p.36]{lewis2012optimal}, namely 
\be\label{eq:Y_infty}
\lim_{N \to \infty} Y = \bY.
\ee
Note that $Y$ and its limit $\bY$ are unknown to Cloud.
We have the following result.
\begin{theorem}\label{thrm:HF_knwon}
	Let Assumption \ref{Assum:sys_stable} hold, and suppose that Cloud knows the matrices $H$ and $F$ in  \eqref{eq:LQ_merged}.
	If the pair $(A, B^\top \bY )$ with $\bY$ in \eqref{eq:Y_infty} is observable,  then Cloud recovers the system matrices $A$, $B$ in \eqref{eq:lin_sys} and the matrix $R$ in \eqref{eq:LQR_cost}.
		Furthermore, from the obtained set of matrices, Cloud infers the  cost matrices $Q$ and $P$ if and only if the set given in \eqref{eq:J_theorem} is a singleton. 
\end{theorem}
\emph{Proof}. See  Appendix \ref{app:proofs}.$\qed$
\mypar
We address next the case when Cloud has only received the true value of $H$ and $x_0^\top F$ in \eqref{eq:LQ_merged}. 
\begin{theorem}\label{thrm:Hx0F_knwon}
	Let Assumptions \ref{Assum: multiple_MPC} and \ref{Assum:sys_stable} hold, and suppose that Cloud knows the matrices $H$ and $x_0^\top F$ in  \eqref{eq:LQ_merged}.
	If $\rank X_0 = n$ and the pair $(A, B^\top \bY)$ is observable where $X_0$ is given in \eqref{eq:all_init} and $\bY$ in \eqref{eq:Y_infty}, then Cloud infers the system matrices $A$ and $B$, and the cost matrices $R$, $Q$, and $P$ as in Theorem \ref{thrm:GWOx0HFx0}.
\end{theorem}
\emph{Proof}.
See  Appendix \ref{app:proofs}.		$\qed$ 
\mypar
Now that we have established the results for inherent privacy guarantees of the dense form of MPC in various side-knowledge scenarios, we shift our attention to adopting random affine transformations in \eqref{eq:LQ_merged}, consistent with Problem \ref{q:main_question}. 
\section{Privacy analysis of affine transformation for MPC in dense form}\label{sec:affine_tra_com}
 We study privacy preserving properties of random affine transformation when it is applied to the dense form of MPC given in \eqref{eq:LQ_merged}. 
\subsection{Affine transformation $(\bR, \br)$}
	Recall the program \eqref{eq:LQ_merged} for obtaining $z^*$
	\be\label{eq:com_form_without_const}
\begin{aligned}
\min_{z} \quad &\frac{1}{2}z^{\top} H z + x_0^{\top}Fz\\
\quad \text{s.t.} \quad & Gz \leq W + Ox_0,
\end{aligned}
\ee
where we have dropped the constant term $x_0^\top Y x_0$ from the cost function \eqref{eq:cost_LQ_merged}.
Plant uses the random affine transformation
\be\label{eq:gen_aff}
z = \bR \zeta + \br,
\ee
as its privacy preserving mechanism where $\bR \in \R^{Nm \times Nm}$ is a randomly chosen invertible matrix and $\br \in \R^{Nm}$ is a randomly chosen vector. The program \eqref{eq:com_form_without_const} under \eqref{eq:gen_aff} is transformed to
\be \label{eq:comp_trans}
\begin{aligned}
	\min_{\zeta} \quad &\frac{1}{2} \zeta^{\top} \tilde{H} \zeta + \tilde{f}^\top \zeta\\
	\quad \text{s.t.} \quad & \tilde{G} \zeta \leq \tilde{e},
\end{aligned}
\ee
where $\tilde{H}  \coloneqq  \bR^ \top H \bR $, $\tilde f  \coloneqq  \bR^\top (F^{\top} x_0  + H \br) $, $\tilde{G}  \coloneqq  G\bR$, and $\tilde{e}  \coloneqq  (W + Ox_0 - G\br)$. 
Note that Plant outsources the computation of the  optimizer $\zeta^*$  for the transformed program \eqref{eq:comp_trans} to Cloud. Clearly, by receiving $\zeta^*$ Plant recovers $z^*$ (the optimizer of \eqref{eq:com_form_without_const}) since the mechanism \eqref{eq:gen_aff} is accurate. 
\mypar 
The following proposition shows what Cloud infers about the key $(\bR, \br)$ of the transformation.
\begin{proposition}\label{prp:key_rev}
	Suppose that Cloud knows the matrices $\tilde{G}$ and $\tilde{e}$ in \eqref{eq:comp_trans}. 
	Cloud infers the transformation parameter as $(\hat\bR, \hat \br) = (\bR, (L \otimes I_m)\br)$ where $L = I_{N} - (\frac{1}{N})\bo_{N}\bo_{N}^\top$.
\end{proposition}
\emph{Proof}. See  Appendix \ref{app:proofs}.$\qed$
\mypar
It follows from Proposition \ref{prp:key_rev} that Cloud infers the key for the privacy preserving mechanism, which enables it to  remove the effect of the mechanism and later on to use the results in Lemma \ref{lem:G_known} and Theorem \ref{thrm:GWOx0HFx0} for obtaining sensitive parameters. This motivates Plant to hide the constraints from Cloud. Notably, Plant can permute the constraints as 
\begin{align*}
\tilde{G}  \coloneqq  \bP G,\, \tilde{e}  \coloneqq  \bP (W+Ox_0),
\end{align*}
where $\mathbf{P}$ is a permutation matrix
whose rows are uniform random rearrangement of the rows of the identity matrix of size $2N(m+p)$. 
The resulting privacy guarantees in this case will be discussed next.

\subsection{Affine transformation and permutation $(\bR, \br, \bP)$}\label{sub:R_r_p}
Let Plant first apply the transformation \eqref{eq:gen_aff} and then the permutation matrix $\bP$  for the constraints as its privacy mechanism. It is easy to verify that the transformed program \eqref{eq:comp_trans} has the following parameters:
\be\label{eq:gen_trans_param}
\begin{aligned}
	\tilde{H} &= \bR^ \top H \bR, \qquad  \tilde{f} = \bR^\top (F^{\top} x_0  + H \br) \\ \tilde{G} &= \bP G\bR, \qquad \tilde{e} = \bP(W + Ox_0 - G\br). 
\end{aligned}
\ee
Note that the above quantities are independent of the order in which Plant applies the transformation \eqref{eq:gen_aff} and the permutation $\bP$. 
Let Assumption \ref{Assum: multiple_MPC} hold. Then, for each $i \in \calI = \{ 1, 2, \ldots, I\}$, Cloud can form the following equations
\bse\label{eq:pair_data_instance_i_F}
\begin{align}
\tilde{f}_i & = \bR^\top (F^{\top} x_{0|i}  + H \br)\label{eq:tilde_fi} \\
z_i^* & = \bR \zeta_i^* + \br, \label{eq:zi_star_F}
\end{align}
\ese
where $z_i^*$ and $\zeta_i^*$ are the optimizer of the original \eqref{eq:com_form_without_const} and the transformed program \eqref{eq:comp_trans}, for $x_0\equiv x_{0|i}$, respectively.
Note that Cloud receives the value $\tilde{f}_i$ \eqref{eq:tilde_fi} from Plant and obtains  $\zeta_i^*$ \eqref{eq:zi_star_F} after solving the optimization. By using \eqref{eq:pair_data_instance_i_F}, Cloud writes
\be\label{eq:tilde_F_system_cloud}
\begin{aligned}
	\tilde{f}_{i+1} & = \bR^\top (F^{\top} x_{0|i+1}  + H \br) \\
	& = \bR^\top F^{\top} Ax_{0|i} + \bR^\top F^{\top}  \bar{B}z_i^{*}  + \bR^\top H \br,
\end{aligned}
\ee
where we have substituted the system dynamics and denoted  $\bar{B}  \coloneqq  [B \quad 0] \in \R^{n \times Nm}$. Furthermore, by using \eqref{eq:pair_data_instance_i_F} and assuming that the matrix $F \in \R^{n \times Nm}$ has full row rank,  it follows that \eqref{eq:tilde_F_system_cloud} can be rewritten as
\be\label{eq:tilde_F_system_cloud_final}
\begin{aligned}
	\tilde{f}_{i+1} & = \bR^\top F^{\top} A(F^\top)^\dagger (\bR^{-\top} \tilde{f}_i -H\br)  \\
	&  \qquad   + \bR^\top F^{\top}  \bar{B}( \bR \zeta_i^* + \br)  + \bR^\top H \br \\
	& = \bR^\top F^{\top} A(F^\top)^\dagger \bR^{-\top} \tilde{f}_i + \bR^\top F^{\top}  \bar{B}\bR \zeta_i^*  + d, 
\end{aligned}
\ee
where 
$$d  \coloneqq  - \bR^\top F^{\top} A(F^\top)^\dagger H\br  +  \bR^\top F^{\top} \bar{B} \br + \bR^\top H\br.$$ To proceed further, for each $i=1, \ldots, I-1$ define 
\be\label{eq:delta_f_delta_zeta}
\begin{aligned}
	\delta\tilde{f}_i   \coloneqq  \tilde{f}_{i+1} - \tilde{f}_{i},  \qquad 
	\delta\zeta_i^*   \coloneqq  \zeta_{i+1}^* - \zeta_{i}^*,
\end{aligned}
\ee
and rewrite the dynamics \eqref{eq:tilde_F_system_cloud_final}  as
\be\label{eq:delta_system}
\delta\tilde{f}_{i+1} = \tilde{A} \delta\tilde{f}_i + \tilde{B} \delta\zeta_i^*,
\ee
where $\tilde{A} \coloneqq   \bR^\top F^{\top} A(F^\top)^\dagger \bR^{-\top} $ and $\tilde{B} \coloneqq  \bR^\top F^{\top}  \bar{B}\bR$. 
For further simplification, denote the data matrices 
\be\label{eq:data_matrices_F}
\begin{aligned}
	\Delta\tilde{F}_0 & \coloneqq  \bbm \delta\tilde{f}_1 & \cdots & \delta\tilde{f}_{I-2} \ebm \in \R^{Nm \times (I-2)} \\
	\Delta\tilde{F}_1 & \coloneqq  \bbm \delta\tilde{f}_2 & \cdots & \delta\tilde{f}_{I-1} \ebm \in \R^{Nm \times (I-2)} \\
	\Delta\tilde{Z}^* & \coloneqq  \bbm  \delta\zeta_1^* & \cdots & \delta\zeta_{I-2}^*\ebm \in \R^{Nm \times (I-2)}.
\end{aligned}
\ee
Note that these matrices are known to Cloud.
Bearing in mind Assumption \ref{Assum: multiple_MPC}, the following 
holds:
\be\label{eq:tilde_e_system_Instances_F}
\Delta\tilde{F}_1 = \tilde{A}\Delta\tilde{F}_0 + \tilde{B}\Delta\tilde{Z}^*,
\ee
where $\Delta\tilde{F}_1$, $\Delta\tilde{F}_0$ and $\Delta\tilde{Z}^*$ are given in \eqref{eq:data_matrices_F}. 
\mypar
We have the following result:
\begin{theorem}\label{thrm:iden_R_r_P_with_F}
 Let Assumption  \ref{Assum: multiple_MPC} hold and suppose that  Cloud knows the transformed program \eqref{eq:comp_trans} with $\tilde{H}$, $\tilde{f}$, $\tilde{G}$ and $\tilde{e}$ given in \eqref{eq:gen_trans_param}. Assume that  $\rank F = n$ and   $\rank \Delta\tilde{F}_0 = n$ 	with $F$ given in \eqref{eq:cost_LQ_merged} and $\Delta\tilde{F}_0$ in \eqref{eq:data_matrices_F}.
 Let $\Delta\tilde{F}_0$ be decomposed as
 \be\label{eq:E0_data}
 \Delta\tilde{F}_0 = {\Delta F_b} E_0,
 \ee
  where ${\Delta F_b} \in \R^{Nm \times n}$ and $E_0 \in \R^{n \times (I-2)}$ are full column rank and full row rank matrices, respectively.\footnote{Note that such a decomposition is always feasible due to $\rank\Delta\tilde{F}_0=n.$}
  Then Cloud identifies $\hat{A}  \coloneqq  TAT^{-1}$ with an unknown invertible matrix $T \in \R^{n\times n}$ if there exists $\Theta \in \R^{(I-2) \times n}$ such that
	\bse\label{eq:rank_cond_data}
	\begin{align}
		E_0 \Theta &= I_n \\
		 \Delta\tilde{Z}^* \Theta &= 0 \label{eq:DeltaZ_cond},
	\end{align}
	\ese
	with $\Delta\tilde{Z}^*$ in \eqref{eq:data_matrices_F}.
\end{theorem}
\emph{Proof}. See  Appendix \ref{app:proofs}.$\qed$
\mypar
It follows from Theorem \ref{thrm:iden_R_r_P_with_F} that under suitable conditions Cloud is able to identify the spectrum of the matrix $A$ in \eqref{eq:lin_sys} even if Plant is using the transformation triple $(\bR, \br, \bP)$ as its privacy mechanism.
We conclude this section with a discussion on how the result of Theorem \ref{thrm:iden_R_r_P_with_F} can be stated for identifying $T\bar{B}\bR$ in \eqref{eq:system_data_final} and how the case of time-varying affine transformations can be handled.

\begin{remark}
	It should be noted that Theorem \ref{thrm:iden_R_r_P_with_F} emphasizes the condition under which the eigenvalues of the system can be identified (through $TAT^{-1}$) in the light of their privacy-sensitive values. The condition \eqref{eq:rank_cond_data} can be extended for also identifying $T\bar{B}\bR$ in \eqref{eq:system_data_final}, which includes $\bR$ the key of the transformation \eqref{eq:gen_aff}  and also the $B$ matrix of the system \eqref{eq:lin_sys}. For this case we need to replace the condition \eqref{eq:rank_cond_data} with the existence of a matrix $\Theta_e \in \R^{(I-2) \times (n+Nm)}$ such that
	\be\label{eq:A_B_gen_system}
	\bbm E_0 \\ \Delta\tilde{Z}^* \ebm \Theta_e = \bdiag(I_n, I_{Nm}),
	\ee
 which is equivalent to the matrix on the left-hand side of above to be full row rank. 
	It is worth mentioning that if $\delta\zeta_i^{*}$ for $i \in \calI = \{ 1, 2, \ldots, I-2\}$ is persistently exciting of order $n+1$, then the mentioned full row rank condition holds \cite{willems2005note}.
	Nonetheless, for cost function \eqref{eq:LQR_cost} with large prediction horizon $N$, the condition \eqref{eq:A_B_gen_system} can be conservative. This is specially true since the input sequence $\{\delta\zeta_i^*\}_{i=1}^{I-2}$ are constrained to be the solution of the MPC problem \eqref{eq:comp_trans}.
	$\qedwhite$ 
\end{remark}

\begin{remark}[Time-varying transformation $(\bR_i,\br_i,\bP_i)$]
	Plant can adopt $(\bR_i, \br_i, \bP_i)$, i.e. a time-varying transformation, for preserving privacy of the parameters in the program \eqref{eq:com_form_without_const}, meaning for each instance $i \in \calI = \{1, 2, \cdots, I\}$ of the program, Plant uses a different affine transformation. Two points need to be raised in this case: First, the results in Theorem \ref{thrm:iden_R_r_P_with_F} also hold for time-varying permutation $\bP_i$. Second:
applying time varying affine transformation $(\bR_i, \br_i)$ can lead to a computationally inefficient program unsuitable for real-time applications of MPC. This is specially noticeable since solving MPC in the dense least square already put some extra local computational burden on Plant.
 It is worth mentioning that the results can be applied to the ``sufficiently slow" time-varying transformation, that is, when $(\bR_{(\cdot)}, \br_{(\cdot)})$ remains constant long enough until the conditions in Theorem \ref{thrm:iden_R_r_P_with_F} are satisfied. $\qedwhite$ 
\end{remark}

\section{Numerical Simulation}\label{sec:num_sim}
In this section, a case study is presented to further examine the results.
\subsection{Case study setup}
\begin{figure}[ht]
	\begin{center}
		\includegraphics[width=0.440\textwidth]{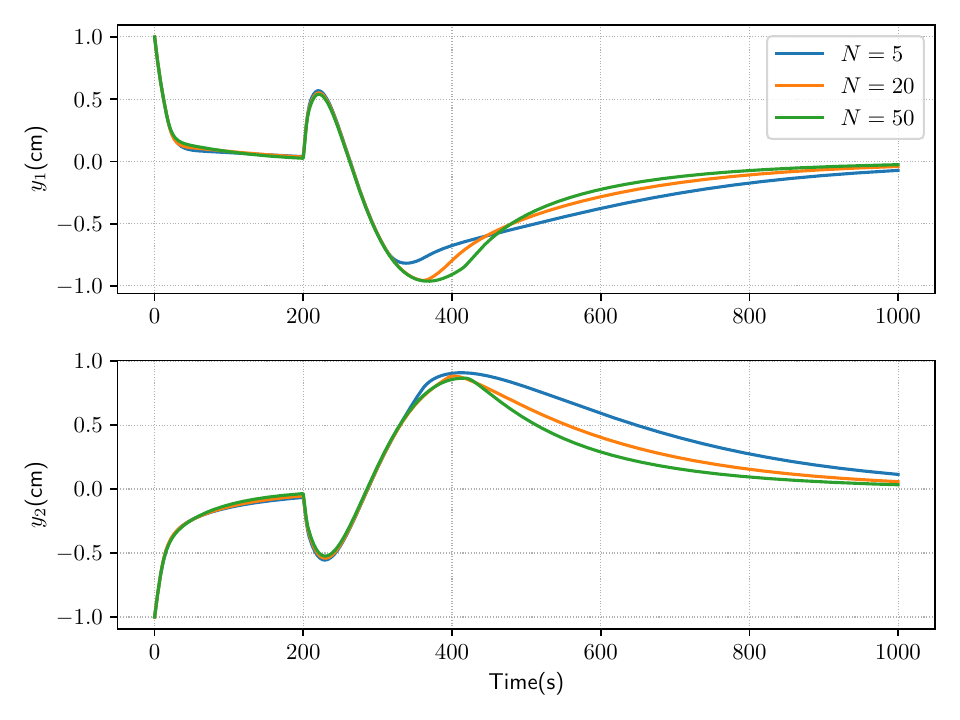}
		\caption{The output $y_1$ and $y_2$ for the system \eqref{eq:Quad_tank_model} for $N=5$, $20$, $50$. The system response is more sluggish with $N=5$ compared to $N=20$ and $N=50$.}\label{fig:outputs}
	\end{center}
\end{figure}

 We use the Quadruple-Tank Process (QTP) as the system to be controlled \cite{johansson2000quadruple}. The QTP consists of four tanks containing water, two valves and two pumps as actuators which change the water levels in the tanks. The QTP is a nonlinear system but its linearized model is
\be\label{eq:Quad_tank_model}
\begin{aligned}
	\dot{x}_1 & = -1/T_1 x_1 + A_3/(A_1T_3) x_3 + \gamma_1k_1/A_1 u_1 \\
	\dot{x}_2 & = -1/T_2 x_2 + A_4/(A_2T_4)x_4 + \gamma_2k_2/A_2u_2 \\
	\dot{x}_3 & = -1/T_3x_3 + (1-\gamma_2)k_2/A_3 u_2 \\
	\dot{x}_4 & = -1/T_4x_4 + (1-\gamma_1)k_1/A_4u_1\\
	y & = \bbm k_cx_1 \\ k_cx_2\ebm,
\end{aligned}
\ee
where $x_i$ , $T_i$ and $A_i$ are the water level, time constant and cross-section for tank $i=1,2,3, 4$, and  $\gamma_j$, $k_j$, and $u_j$ are constants for valve, pump and the voltage applied to the pump $j = 1, 2$, and $k_c$ is the sensor coefficient. The cross-sections for tanks are $ (A_1 , A_2, A_3, A_4) =(28, 32, 28, 32)\text{cm}^2$, and $k_c = 0.5$.
We use the values 
$(T_1, T_2,T_3, T_4) = (63, 91, 39, 56)$, and $(k_1, k_2, \gamma_1, \gamma_2) = (3.14, 3.29, 0.43, 0.34)$ given in \cite{johansson2000quadruple} for the operating
point and discretize the system using zero-order hold method with the sampling rate $T_s = 2s$. For the later reference (Subsections \ref{sub:HF_sim}-\ref{sub:eigen_value_estimation}) we note the spectrum of discretized system matrix $\spec(A)= \{0.968, 0.978, 0.950, 0.964\}$.
The control objective is to reject the process disturbance $d_k = \bbm 3 & -3 \ebm^\top (0.99)^{k-200}$ for $k \geq 200$ while the system is at rest and the constraints $ -2 \leq y_i(k) \leq 2$ and $ -1 \leq u_i(k) \leq 1$ for $i = 1, 2$ are satisfied. For achieving this we use $Q = 2I_4$, $R=I_2$ and $P=0$ for the cost matrices in \eqref{eq:LQR_cost} and use the dense form of MPC given in \eqref{eq:LQ_merged}\footnote{Simulation files are available at \url{https://github.com/teimour-halizadeh/affine_transformation_code}}. 
\mypar 
\textit{Privacy concerns:} One of the major concerns from Plant's perspective is data privacy during an outsourced search for optimal control inputs.
Notable in this case, the system in the selected operating point has a non-minimum phase zero at $s= 0.013$ or out of unit-disc zero in discrete time at $z=1.02$. 
This non-minimum phase zero turns the system \eqref{eq:Quad_tank_model} into a possible target for zero dynamics attack, where the attacker can cause instability for the system by injecting false data \cite{teixeira2015secure}.
On the other hand, as it has been studied in detail in \cite[p.170-175]{camacho2007model}, the zero for this system can deteriorate the performance of MPC, and even instability of system can arise when the controller attempts to cancel the non-minimum phase zero with right hand side poles. To obtain a satisfactory performance for this system (and generally for system with non-minimum phase zeros) the prediction horizon parameter $N$ must be selected sufficiently large.  Therefore, Plant needs to take into account in its design the effects of prediction horizon $N$ and preserving privacy of the dynamics \eqref{eq:Quad_tank_model} from Cloud. 
\mypar
Figure \ref{fig:outputs} shows the output of the linearized system using dense MPC with prediction horizon $N= 5, 20$ and $50$  and total time steps $K=500$. As it can be seen from the figure, the controller performs better when $N = 50$ compared to $N=5$. Furthermore, the root mean squared error $e_{N} = (\norm{y}_2/K^{1/2})$ for the three scenarios are $e_{5} = 0.67$, $e_{20} = 0.55$, and  $e_{50} = 0.49$ which further validates the superior performance for $N = 50$. Therefore, Plant prefers to adopt $N = 20$ or $N=50$ rather than $N=5$ as the prediction horizon for MPC. 
\subsection{Estimation using $H$ and $F$}\label{sub:HF_sim}
To examine how Cloud can identify the system dynamics using Theorem \ref{thrm:HF_knwon} (by \textit{only} accessing $H$ and $F$ in \eqref{eq:LQ_merged}) we have computed the $\bar{Y}$ in \eqref{eq:Y_infty} (which is unknown to Cloud) since it satisfies the Lyapunov equation $A^\top \bar{Y}A - \bar{Y} = -Q$ \cite[p.37]{lewis2012optimal}.
Furthermore, from the spectrum of $A$ we can see that the system is stable, and the pair $(A, B^\top \bY )$ is observable. Therefore, all the conditions of Theorem \ref{thrm:HF_knwon} are satisfied, and hence Cloud can use the theorem to infer the system and cost matrices. 
To measure the error for Cloud's estimation of the system matrix $A$ we define the error index
\be\label{eq:A_error_index}
\epsilon_A = \frac{\norm{A-\hat{A}}_{F}}{\norm{A}_{F}},
\ee
where $\norm{\cdot}_{F}$ is Frobenius norm. For matrix $B$ an analogues index to  \eqref{eq:A_error_index} will be used, denoted as $\epsilon_B$. In addition, we have shown the error between $Y = \sum_{i=0}^{N-1}{(A^{\top})^{i}QA^{i}} + (A^{\top})^{N}PA^{N}$ given in \eqref{eq:LQ_merged} and $\bar{Y}$ in \eqref{eq:Y_infty} as $N$ changes by computing $\epsilon_Y = {\norm{\bar{Y}-Y}_{F}}/{\norm{\bar{Y}}_{F}}$. With regard to $Q$ and $P$ matrices, as we argued in Theorem \ref{thrm:GOHF_known} and Remark \ref{rem:P_Q_unique} for the choice $Q \succ 0$ and $P \succeq 0$ with $B$ in the system \eqref{eq:Quad_tank_model} the set \eqref{eq:J_theorem} is not a singleton and hence Cloud cannot obtain unique values for $P$ and $Q$ but only the uncertainty set \eqref{eq:J_theorem}. 
\begin{figure}[ht]
	\begin{center}
		\includegraphics[width=0.440\textwidth]{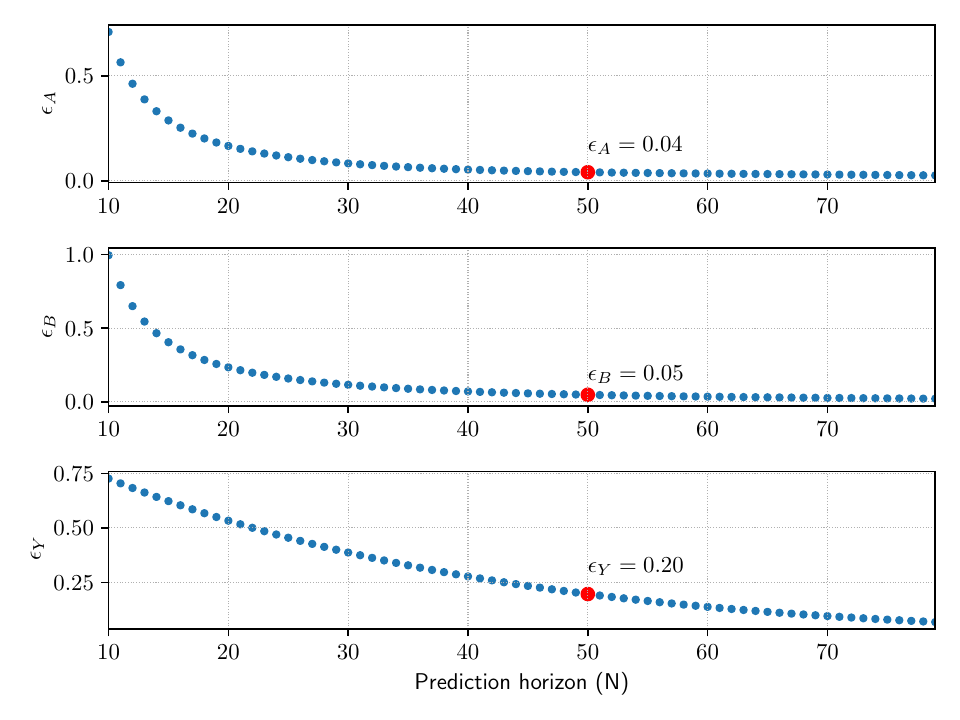}
		\caption{Cloud's estimation error for $A$ and $B$ matrices of quadruple-tank process, and convergence error for $Y$ using Theorem \ref{thrm:HF_knwon}}\label{fig:Alg_result}
	\end{center}
\end{figure}
\begin{figure}[ht]
	\begin{center}
		\includegraphics[width=0.440\textwidth]{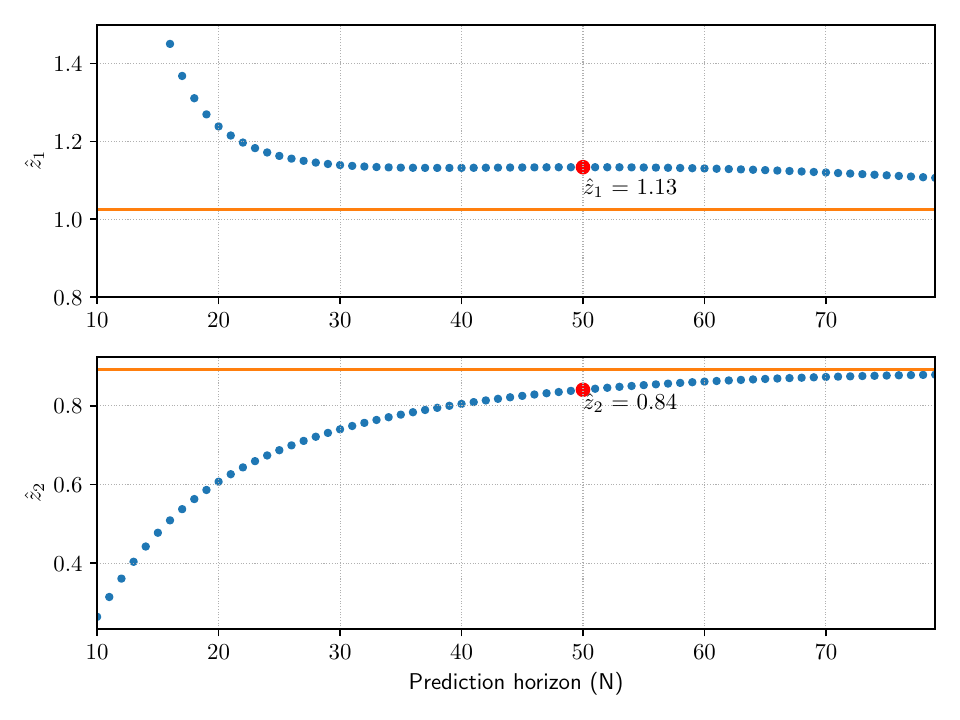}
		\caption{Cloud's estimation of the true zeros of the system $z_1 = 1.02$, and $z_2 = 0.89$ as prediction horizon $N$ changes }\label{fig:zero_location}
	\end{center}
\end{figure}
As Figure \ref{fig:Alg_result} shows Cloud is able to identify $A$ and $B$ matrices with less than $5 \%$ error when $N = 50$, that is in order for Cloud to apply Theorem \ref{thrm:HF_knwon}, it needs not $N \to \infty$. We have also included  in Figure \ref{fig:zero_location} what Cloud infers for the zeros of the system $z_1 = 1.02$, and $z_2 = 0.89$ using its  estimated $\hat{A}$, $\hat{B}$ and $C$. As we can observe from this figure, Cloud's estimation improves as Plant adopts higher values for $N$ to the degree that  at $N=50$ it can estimate both zeros with less than $10 \%$ error. Overall, Cloud's estimations can be considered as a privacy breach  for Plant.
\subsection{Affine transformation and permutation}\label{sub:eigen_value_estimation}
\begin{figure}[ht]
	\begin{center}
		\includegraphics[width=0.440\textwidth]{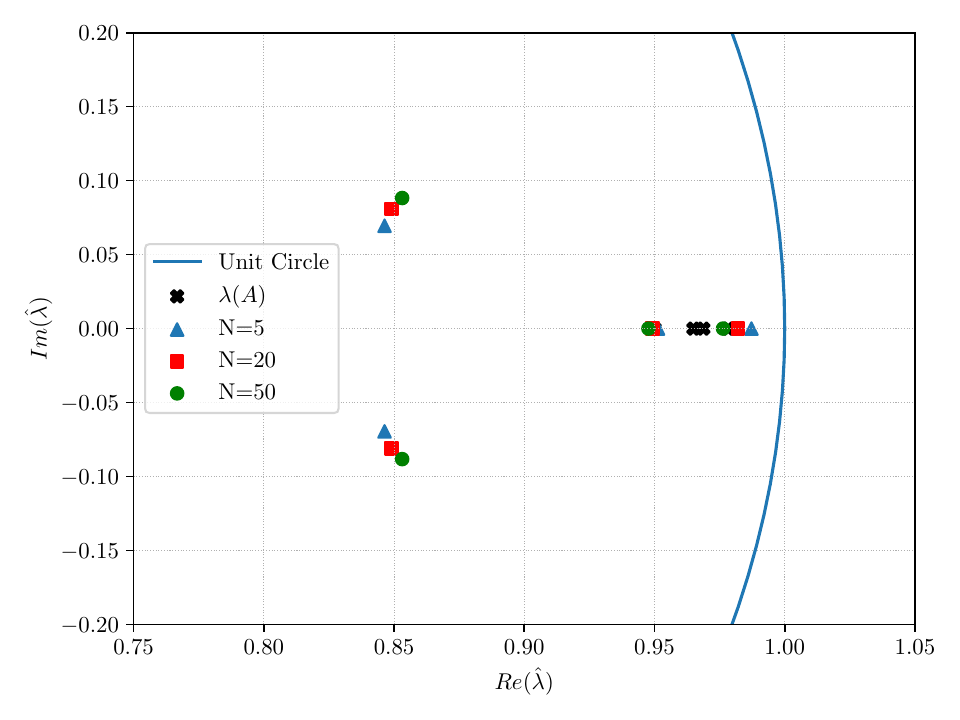}
		\caption{Cloud's estimation of eigenvalues of $\hat{A} = TAT^{-1}$ for $N=5$, $20$, $50$ using Theorem \ref{thrm:iden_R_r_P_with_F} when Plant has adopted random $(\bR, \br, \bP)$ and true values of $\lambda(A)$.}\label{fig:eigen_values}
	\end{center}
\end{figure}

As we  studied in Section \ref{sec:affine_tra_com}, Plant can adopt random affine transformation and permutation $(\bR, \br, \bP)$ to prevent Cloud from accessing sensitive information (See Figure \ref{fig:Alg_result} and \ref{fig:zero_location}). For the matrices $\bR$ and $\br$ given in \eqref{eq:gen_aff} we consider
$\bR_{i,j}, \br_{i} \in [-10^{3}, 10^{3}]$, i.e., each element of $\bR$ and $\br$ is selected uniformly randomly from the given interval. By adopting this mechanism, Plant outsources the computation of the optimizer for transformed program \eqref{eq:comp_trans} with matrices \eqref{eq:gen_trans_param} to Cloud. Since the mechanism is exact, the system \eqref{eq:Quad_tank_model} controlled with solving \eqref{eq:comp_trans} has the same responses as in Figure \ref{fig:outputs} for $N=5, 20$, and $50$; hence we do not present them again.
\mypar
From the system and cost matrices, it can be checked that $\rank F = 4$. Furthermore, from the obtained data in \eqref{eq:tilde_fi} it holds that $\rank \Delta\tilde{F}_0 = 4 $ with $\Delta\tilde{F}_0$ in \eqref{eq:data_matrices_F}. Therefore, Cloud can apply Theorem \ref{thrm:iden_R_r_P_with_F} for estimating eigenvalues of $A$.
It should be noted that the condition $\Delta\tilde{Z}^* \Theta = 0$ in \eqref{eq:DeltaZ_cond} has been relaxed to $\norm{\Delta\tilde{Z}^* \Theta}_{F} \leq \epsilon$ where the value $\epsilon$ is minimized. We have shown the results of eigenvalue estimation by Cloud in Figure \ref{fig:eigen_values}.
As it can be seen from this figure, Cloud is able to approximately estimate the eigenvalues of $A$. While these estimations can indeed be considered as privacy leakage of the random affine transformation method, we need to point out that since the condition \eqref{eq:DeltaZ_cond} was replaced by a surrogate and the true eigenvalues cannot be exactly recovered.

\section{Conclusion}\label{sec:con}
We have studied privacy preserving properties of random affine transformation that are applied to computationally demanding MPC problems before outsourcing to Cloud the search for the optimizer. We have shown that in two forms of MPC, namely separate and dense form, these transformations are vulnerable to mild side-knowledge from Cloud. In other words, while they can be used to create some ambiguity for a privacy-sensitive value in the form of an uncertainty set (as it has been done so in the literature), it is not necessarily correct to argue that they \textit{guarantee} the same privacy degree under side-knowledge from Cloud which its amount cannot be  predetermined. Future directions for this research include analyzing the privacy preserving properties of time-varying transformations, studying banded least square form of MPC and floating point analyses of random variables.

\ack
Nils Schl{\"u}ter and Moritz Schulze Darup acknowledge 
financial support by the German Research Foundation (DFG) under the grant SCHU 2940/4-1.

\appendix
\section{Proofs}\label{app:proofs}
\emph{Proof of Theorem \ref{thrm:LQ_sep_main}}.
	Recall that Cloud receives $(\tilde{A}, \tilde{B}, \tilde{C})$ and $\tilde{J}(\bxt, \but)$ given in \eqref{eq:lin_transformed} and \eqref{eq:mod_cost} from  Plant.
 Note the expanded form for $\tilde{M}$ in \eqref{eq:mod_cost} is
	\be\label{eq_M_tilde}
	\begin{aligned}
		\tilde{M} = \bbm \tilde{M}_{1,1} & \star \\ -(\bG^{-\top}R\bG^{-1})\bF\bT^{-1} & \bG^{-\top}R\bG^{-1} \ebm,
	\end{aligned}
	\ee
	with 
	$
	\tilde{M}_{1,1} = \bT^{-\top} Q\bT^{-1} + (\bF\bT^{-1})^{\top}(\bG^{-\top}R\bG^{-1})(\bF\bT^{-1})
	$, and $(\star)$ denotes the transpose of block $\tilde{M}_{2,1}$.
The following set of observations prove the claim:	
	\begin{enumerate}
		\item Cloud finds the non-singular matrix $\bG^{-\top}R\bG^{-1}$ from block $(2,2)$ in $\tilde{M}$.
		\item From block $(2,1)$ in $\tilde{M}$ and using the known matrix $\bG^{-\top}R\bG^{-1}$, Cloud infers the value of the matrix $\bF \bT^{-1}$. 
		\item From block $(1,1)$ in $\tilde{M}$ and using the previously identified matrices $\bG^{-\top}R\bG^{-1}$ and $\bF \bT^{-1}$, Cloud computes $\bT^{-\top} Q\bT^{-1}$.
		\item From $\tilde{A} = \bT A \bT^{-1}-\bT B \bG^{-1} \bF \bT^{-1}$ and $\tilde{B} = \bT B \bG^{-1}$ (see \eqref{eq:lin_transformed}), Cloud recovers $\tilde{A} = \bT A \bT^{-1} - \tilde{B}(\bF \bT^{-1})$. Thus, the matrix $\bT A \bT^{-1}$ is revealed to Cloud. $\qed$ 	
	\end{enumerate}
 			\emph{Proof of Corollary \ref{cor:affine}}. The proof follows from the observation that the expressions of $\tilde{A}$, $\tilde{B}$ in \eqref{eq:aff_tr_system} are the same as those in \eqref{eq:lin_transformed}. Furthermore, from the  quadratic terms of $\tilde{J}(\bxt, \but)$, which also appear in the transformed cost function \eqref{eq:aff_tr_cost}, 
   Cloud recovers the matrices  $(\bT^{-\top} Q\bT^{-1} + (\bF\bT^{-1})^{\top}(\bG^{-\top}R\bG^{-1})\bF\bT^{-1})$,
   $(\bG^{-\top}R\bG^{-1})\bF\bT^{-1}$ and $\bG^{-\top}R\bG^{-1}$. 
   Note that $\tilde{J}_{r}(\bxt, \but) - \tilde{J}(\bxt, \but)$ does not include any quadratic terms to create uncertainty for Cloud.
			 By using these matrices, Cloud can follow the same steps in the proof of Theorem \ref{thrm:LQ_sep_main} to obtain the same results, i.e., it first finds  $\hat{R}=\bG^{-\top}R\bG^{-1}$ and then $\hat{Q} = \bT^{-\top} Q\bT^{-1}$ and $\hat{A} = \bT A\bT^{-1}$. $\qed$
    \mypar 
    \emph{Proof of Corollary \ref{cor:high_dim_sep}}.
	We present the proof by following analogous steps in the proof of Theorem \ref{thrm:LQ_sep_main}.
 	Cloud receives $(\tilde{A}, \tilde{B}, \tilde{C})$ and $\tilde{J}_{h}(\tilde{\bx}, \tilde{\bu})$ given in \eqref{eq:sys_T_high} and \eqref{eq:cost_T_high} from  Plant.
\mypar
		(1) Cloud finds $(\cbG^{\ell})^{\top}R\cbG^{\ell}$ from block $(2,2)$ of $\tilde{M}$.
		(2) From block $(2,1)$ in $\tilde{M}$ and using the obtained matrix $(\cbG^{\ell})^{\top}R\cbG^{\ell}$, Cloud infers the following equality 
concerning $\cbF \cbT^{\ell} \in \R^{\bar{m}\times \bar{n}}$:
		\be\label{eq:FTL}
		\cbF \cbT^{\ell} = (\cbF \cbT^{\ell})_{\text{p}} + X,
		\ee
		where $(\cbF \cbT^{\ell})_{\text{p}}$ is any particular solution to $- \tilde{M}_{2,1} = ((\cbG^{\ell})^{\top}R\cbG^{\ell})(\cbF \cbT^{\ell})_{\text{p}}$ 
		and $X \in \R^{\bar{m}\times \bar{n}}$ is \textit{any} matrix satisfying 
		$((\cbG^{\ell})^{\top}R\cbG^{\ell}) X = 0$. As $R \succ 0$, the latter is equivalent to $X\in \ker \cbG^{\ell}$.
	Note that, by construction, a particular solution always exists. 
		(3) By substituting the known term $(\cbG^{\ell})^{\top}R\cbG^{\ell}$ and the expression of $\cbF \cbT^{\ell}$ from \eqref{eq:FTL} into $\tilde{M}_{1,1}$,  Cloud finds the matrix $(\cbT^{\ell})^{\top} Q\cbT^{\ell}$ following the derivations below:
\begin{align*}
&\tilde{M}_{1,1}   =
   ((\cbF \cbT^{\ell})_{\text{p}} + X)^{\top}(\cbG^{\ell})^{\top}R\cbG^{\ell}((\cbF \cbT^{\ell})_{\text{p}} + X) \\
  &  + (\cbT^{\ell})^{\top} Q\cbT^{\ell}   
 = (\cbT^{\ell})^{\top} Q\cbT^{\ell} + (\cbF \cbT^{\ell})_{\text{p}} ^{\top}(\cbG^{\ell})^{\top}R\cbG^{\ell}(\cbF \cbT^{\ell})_{\text{p}},
\end{align*}
hence
$
 (\cbT^{\ell})^{\top} Q\cbT^{\ell}  = 
\tilde{M}_{1,1} - (\cbF \cbT^{\ell})_{\text{p}}^{\top}((\cbG^{\ell})^{\top}R\cbG^{\ell})(\cbF \cbT^{\ell})_{\text{p}}
$.
\mypar
		(4) From $\tilde{A} = \cbT A\cbT^{\ell} - (\cbT B\cbG^{\ell})\cbF\cbT^{\ell}$ and $\tilde{B} = \cbT B\cbG^{\ell}$, Cloud obtains $\tilde{A} =\cbT A\cbT^{\ell} - \tilde{B}(\cbF\cbT^{\ell})$. Thus, by substituting $\cbF\cbT^{\ell}$ from \eqref{eq:FTL} into $\tilde{A}$, it follows that $\cbT A \cbT^{\ell}$ is revealed to Cloud.
\mypar
	The proof for the second part of Corollary follows from the fact that for two arbitrary matrices $X_{1} \in \R^{q\times r}$ and $X_{2}\in \R^{r \times q}$  we have $\mspec_{nz}(X_{1}X_{2}) = \mspec_{nz}(X_{2}X_{1})$\cite[Prop. 6.4.10]{bernstein2018scalar}. By considering $X_{1} = \cbT$ and $X_{2} = A\cbT^{\ell}$, we conclude that  
	$
	\mspec_{nz}(\cbT A \cbT^{\ell}) = \mspec_{nz}(A)$,
	which completes the proof. $\qed$ 
    \mypar
    \emph{Proof of Theorem \ref{thrm:GOHF_known}}. We present the proof for $N = n+1$. By using ${\calO}$ in \eqref{eq:constraint_all_block_matrices} and block partitioning it, the following equality holds:
\be\label{eq:su_known_1}
\underbrace{\col(\calO_{1,1}, \cdots, \calO_{1,n})}_{\eqqcolon{\calO}^-} A = \col(\calO_{1,2}, \cdots, \calO_{1,n+1}).
\ee
Notice $\rank {\calO}^- = n$, i.e., ${\calO}^-$ is full column rank since $\calO^{-}$ can be written as
$
\calO^{-} = \col(C, CA, \cdots, CA^{n-1})A,
$
which is the observability matrix 
multiplied by an invertible matrix $A$ (Assumption \ref{Assum: A_inv}). As the pair $(A, C)$ is observable, we find that $\calO^-$ has full column rank.
Hence the state matrix $A$ can be found uniquely from \eqref{eq:su_known_1}.
\mypar
By using the first block in $\calO$, i.e., $CA$, and the known $A$, Cloud infers the output matrix $C$. In addition, Cloud recovers the input matrix $B$ by noticing that the first column block of ${\calG}$ for $N = n+1$ is 
	$$
	\col(C, CA, \cdots, CA^{n}) B = \col( \calG_{1,1}, \calG_{2,1}, \cdots, \calG_{N, 1}),
	$$
which returns a unique value for $B$ due to observability  of $(A, C)$. 	
To recover the matrix $R$, consider $F_{1,N}$ and $H_{N,N}$ blocks from $F$ and $H$ which are
	\bse
	\begin{align}
	2(A^{N})^\top PB &= F_{1,N} \label{eq:F_known} \\
	2(R + B^\top PB) &= H_{N,N} \label{eq:H_known}.
	\end{align}
	\ese
	From \eqref{eq:F_known}, Cloud recovers the matrix $PB$, and then from \eqref{eq:H_known} it computes the  cost matrix $R$. 
\mypar	From the obtained set of matrices, Cloud can form the set
\be\label{eq:set_Q_P_original}
\begin{aligned}
\Big\{ \hat{Q}, \hat{P} \in \R^{n \times n}| &\hat {\calQ}  \coloneqq \bdiag(\hat{Q} , \ldots, \hat{Q} ,\hat{P}) \, \text{and}\, \\
 \bbm \calS & \calT\ebm^\top \hat {\calQ} \calS &= \frac{1}{2} \col(H -2\calR , F) \Big\},
\end{aligned}
\ee
where the matrices $\calR$, ${\calS}$ and $\calT$ are defined in \eqref{eq:cost_all_block_matrices}.  Note that Cloud knows $\calR$, ${\calS}$ and $\calT$ since it has already recovered $A$, $B$, $R$ and received $H$ and $F$.
 Notice that by construction of the received matrices, Cloud can find a pair such as $(\hat{P}\succeq 0,   \hat{Q} \succeq 0)$
	satisfying \eqref{eq:set_Q_P_original}. Any other solution for \eqref{eq:set_Q_P_original} can be written in the following form
	\be\label{eq:gen_P_Q_sol}
	\hat{P} + \Delta P \succeq 0, \quad  \hat{Q} + \Delta Q \succeq 0 ,
	\ee
	where $\Delta P $ and $\Delta Q $ are symmetric.
The pair $(\hat{P} + \Delta P, \hat{Q} + \Delta Q)$ is a solution to \eqref{eq:set_Q_P_original} if and only if $(\Delta P, \Delta Q)$ satisfies
	\be\label{eq:calS_calQ_zero}
	\bbm \calS & \calT\ebm^\top \Delta \calQ \calS= 0,
	\ee
	where $\Delta \calQ = \bdiag(\Delta Q, \cdots,\Delta Q, \Delta P)$. Denote $ \calT^\top \Delta\calQ \calS \eqqcolon\Delta F$ and $ \calS^\top \Delta\calQ \calS \eqqcolon\Delta H$. 
We note that \eqref{eq:calS_calQ_zero} can be rewritten as $\Delta F=0$ and $\Delta H=0$. Consistent with $F$ and $H$, we partition the matrices $\Delta F$ and $\Delta H$.
 We observe that
	\be\label{eq:Delta_F_1}
	\begin{aligned}
		& \Delta F_{1,N}  = (A^{N})^\top \Delta P B
		    = 0  \iff \Delta P B = 0.
	\end{aligned}
	\ee
 Note that the deduction is due to invertibility of $A$.
	Hence, it follows from \eqref{eq:Delta_F_1} that $\Delta P = X_{1}$ for some $X_{1} \in \R^{n \times n}$ with $X_1B = 0$.
By substituting $\Delta P$  in $\Delta F_{1,N-1}$ block, we have
	\be\label{eq:DeltaF_N-1}
	\begin{aligned}
		&\Delta F_{1,N-1} = (A^{N-1})^\top\big(\Delta Q + A^{\top}\Delta PA\big)B \\
		& \qquad \qquad  = 0 \iff \underbrace{(\Delta Q + A^{\top}X_{1}A)}_{\coloneqq X_2} B = 0 ,
	\end{aligned}
	\ee
	where the first line follows from the definition of the block $\Delta F_{1,N-1}$ and the equivalence is due to  $\Delta P = X_1$ and invertibility of $A$.
	Hence, it follows from \eqref{eq:DeltaF_N-1} that $\Delta Q = X_{2} - A^\top X_{1}A $ for some $X_2\in \R^{n \times n} $ with $X_{2}B = 0$.
	Similarly, by substituting $\Delta P$ and $\Delta Q$ in $\Delta F_{1,N-2}$ block it can be verified that
	\be\label{eq:DeltaF_N-2}
		\Delta F_{1,N-2} =  0 \iff 	 (X_{2} -  X_{1})AB = 0.
	\ee
 Analogously, by substituting $\Delta P$ and $\Delta Q$ into  $\Delta F_{1, N-3}$, $\Delta F_{1, N-4}$, $\cdots$, $\Delta F_{1,1}$ it follows:
	\be\label{eq:DeltaF_set}
	\begin{aligned}
		\Delta F_{1,N-3} &= 0 \iff  (X_{2} -  X_{1})A^2B=0 \\
		& \qquad \quad  \vdots\\
		\Delta F_{1,1} &=  0 \iff (X_{2} -  X_{1})A^{N-2}B=0.
	\end{aligned}
	\ee
	Therefore, in order for $\Delta P$ and $\Delta Q$ to satisfy $\Delta F = 0$ in  \eqref{eq:calS_calQ_zero} it follows that we \textit{equivalently} must ensure the existence of 
	$X_{1}$ and $X_{2}$ such that 
	\eqref{eq:Delta_F_1}-\eqref{eq:DeltaF_set} hold.
	Let $X_{1}$ and $X_{2}$ satisfy \eqref{eq:Delta_F_1}-\eqref{eq:DeltaF_set}. It follows that
	\be\label{eq:cont_contradict}
	(X_{1}-X_{2}) \bbm B & AB & \cdots & A^{N-2}B \ebm = 0.
	\ee
 Noting that $N =n+1$ by assumption and that $(A, B)$ is controllable, it follows that  
 $X_{1} = X_{2}\eqqcolon X$ for some $X \in \R^{n \times n}$.  Recalling that $\Delta P = X_1$ and $\Delta Q = X_{2} - A^\top X_{1}A $, the matrix
	$\Delta F$ in \eqref{eq:calS_calQ_zero} is equal to zero if and only if 
	\be\label{eq:DeltaP_DeltaQ}
	\begin{aligned}
		&\Delta P = X  \\
		&\Delta Q = X - A^\top XA, \quad \text{for}\,\, XB = 0.
	\end{aligned}
	\ee 
 To complete the proof, we need to prove that $\Delta P$ and $\Delta Q$ given in \eqref{eq:DeltaP_DeltaQ} satisfy $\Delta H = 0$ in \eqref{eq:calS_calQ_zero}.
	It can be observed that the following relation between the blocks in $\Delta H$ and $\Delta F$ exists:
	\be\label{eq:DeltaH_DeltaF}
	\begin{aligned}
		\bbm 2\Delta H_{1,1} & \star & \cdots & \star \\
		\Delta H_{2,1} & 2\Delta H_{2,2} &  \cdots & \star  \\
		\vdots & \vdots & \ddots & \star  \\
		\Delta H_{N,1} & \Delta H_{N,2} & \cdots & 2\Delta H_{N,N}
		\ebm
		=  \Delta\calF^\top\calA \calS + (\Delta\calF^\top\calA \calS)^\top
	\end{aligned}
	\ee
	where $\Delta\calF \coloneqq \bdiag(\Delta F_{1, 1}, \Delta F_{1, 2}, \cdots, \Delta F_{1, N} )$ and  $\calA \coloneqq \bdiag (A^{-1}, A^{-2}, \cdots, A^{-N})$ and $\calS$ is given in \eqref{eq:cost_all_block_matrices}.
	Thus, from \eqref{eq:DeltaH_DeltaF} it follows that $\Delta H = 0$ if $\Delta\calF = 0$.  
 Therefore, it follows that the pair
	$$
	\hat{P} + X \succeq 0, \quad  \hat{Q} + X - A^\top XA \succeq 0,
	$$
	with $XB = 0$ specifies all the solution for \eqref{eq:set_Q_P_original}.
	Hence, Cloud infers $Q$ and $P$ uniquely, if and only if the set \eqref{eq:J_theorem} is singleton. Note that the proof also holds for $N > n+1$ since the observability and controllability matrices (See \eqref{eq:su_known_1} and \eqref{eq:cont_contradict}), from which the system and cost matrices are obtained, remain full rank.  $\qed$ 
 \mypar 
\emph{Proof of Lemma \ref{lem:G_known}}.
	By having access to $h(1)$, $h(2)$, $\ldots$, $h(N)$ defined in \eqref{eq:markov} with $N = 2n+1$ in the $G$ matrix \eqref{eq:constraint_all_block_matrices}, Cloud obtains the Hankel matrix $\mathcal{H}_{n+1, n+1}$ constructed from Markov parameters, namely 
	$$
	\mathcal{H}_{n+1, n+1} =  \bbm h(1)  & h(2) & \ldots & h(n+1) \\
	h(2) & h(3) & \cdots &  h(n+2) \\
	\vdots & \vdots & \ddots & \vdots \\
	h(n+1) & h(n+2) & \cdots & h(2n+1)
	\ebm.
	$$	
	Then based on Lemma 3.5 in \cite[Lemma 3.5]{verhaegen2007filtering}, Cloud recovers the matrices $A$, $B$, and $C$ up to a similarity transformation $T$, i.e.,
	$
	(A, B, C) = (TA_tT^{-1}, TB_t, C_{t}T^{-1}),
	$
	where $A_t$, $B_t$, and $C_t$ are computed using singular value decomposition (SVD) of $\mathcal{H}_{n+1, n+1}$. Note that the results also hold for $N > 2n+1$ \cite[Lemma 3.6]{verhaegen2007filtering}. $\qed$ 
\mypar
\emph{Proof of Theorem \ref{thrm:GWOx0HFx0}}.
	Since $N\geq 2n+1$ from Lemma \ref{lem:G_known}, Cloud recovers the system matrices as $(A, B, C) = (TA_tT^{-1}, TB_t, C_{t}T^{-1})$. By substituting the obtained $(A,B)$  in ${\calS}$ and ${\calT}$ given in \eqref{eq:cost_all_block_matrices} and defining $V \coloneqq I_N \otimes T$, it follows that
\be\label{eq:Sbar_Tbar_upto}
\begin{aligned}
	&  {\calS}_{t} \coloneqq V^{-1} \calS \\
	& {\calT}_{t} \coloneqq V^{-1} {\calT}  T.
\end{aligned}
\ee
Furthermore, substituting $\calS$ and $\calT$ given in \eqref{eq:Sbar_Tbar_upto} into the $H$ and $F$ matrices in \eqref{eq:cost_LQ_merged} yields
\bse\label{eq:HF_sim}
\begin{align}
&2{\calR} + 2{\calS}_{t}^{\top}{\calQ}_{t}{\calS}_{t} = H \label{eq:H_transformed} \\
&2{T^{-\top}} {\calT}_{t}^{\top} {\calQ}_{t} {\calS}_{t} = F, \label{eq:F_transformed}
\end{align}
\ese
for Cloud where 
$
{\calQ}_{t} \coloneqq V^{\top}{\calQ}V
$.
\mypar 
From the right-hand side of the constraint \eqref{eq:constraint_LQ_merged} the rows $N(2m+p)+1$ to $2N(m+p)$ read as
	\be\label{eq:one_instance}
	-\bo_N \otimes y_{\text{min}}  + {\calO} x_{0}.
	\ee
	Due to Assumption \ref{Assum: multiple_MPC}, Cloud writes $I$ instances of \eqref{eq:one_instance} as $-\bo_N \otimes y_{\text{min}}  + {\calO} x_{0|1} \eqqcolon e_1,$ $\cdots$, $	-\bo_N \otimes y_{\text{min}}  + {\calO} x_{0|I} \eqqcolon e_I$
	which can  be compactly rewritten as
	\be\label{eq:cons_accross_time_compact}
		\big(-\bo_N \otimes y_{\text{min}}\big) \bo_{I}^\top + {\calO} X_0 
		= 	\underbrace{\bbm e_1 & e_2 & \ldots & e_I \ebm}_{\eqqcolon E_0},
	\ee
	with $X_0$ given in \eqref{eq:all_init}. Similarly, Cloud  writes
	\be\label{eq:F_top_X0}
	2({\calT}^{\top}{\calQ}{\calS})^\top X_0 = F^\top X_0
	\ee
	for $I$ instances of $x_0^\top F$.
	By substituting the obtained system matrices $(A, C) = (TA_tT^{-1}, C_tT^{-1})$ into ${\calO}$ at \eqref{eq:cons_accross_time_compact} Cloud has
		\be\label{eq:Eut_final}
	\big(-\bo_N \otimes y_{\text{min}}\big) \bo_{I}^\top + {\calO}_t {T^{-1} X_0} = E_0,
	\ee
	where ${\calO}_t \coloneqq  \col(C_tA_t, C_tA_t^{2}, \cdots, C_tA_t^{N})$. Analogously, substituting $F$ from \eqref{eq:F_transformed} into \eqref{eq:F_top_X0} yields
	\be\label{eq:x0F_accross}
	\begin{aligned}
		2\big({\calT}_{t}^{\top} {\calQ}_{t} {\calS}_{t}\big)^{\top}	T^{-1}X_0= F^\top X_0
	\end{aligned}
	\ee
	for Cloud. Then, Cloud multiplies \eqref{eq:Eut_final} from right by  $L \coloneqq I_I - (\frac{1}{I})\bo_{I}\bo_{I}^\top$ to get
	\be\label{eq:L_for_O}
	{\calO}_t {T^{-1} X_0} L= E_0L,
	\ee
	from which it obtains the unique value $T^{-1}  X_0L = \calO_t^\dagger E_0L$ since ${\calO}_t$ is full column rank and known to Cloud.
	Furthermore, multiplying \eqref{eq:x0F_accross} from right by $L$ yields
	\be\label{eq:L_for_F}
	2\big({\calT}_{t}^{\top} {\calQ}_{t} {\calS}_{t}\big) ^{\top}	T^{-1}X_0L= F^\top X_0L.
	\ee
	Since due to assumption $\rank X_0 = n$ and 
	$\bo_I \notin \im X_0^\top$, 
	equivalently it holds that $\rank X_0L = n$ and subsequently $\rank T^{-1}X_0L = n$. Hence, given that Cloud knows $T^{-1}X_0L$ from \eqref{eq:L_for_O} it determines from \eqref{eq:L_for_F} the unique value $2{\calT}_{t}^{\top} {\calQ}_{t} {\calS}_{t} \eqqcolon F_t$, i.e.,
	\be\label{eq:TtQtSt}
	F_t = (F^\top X_0L)(T^{-1}X_0L)^\dagger.
	\ee
	Then from block $(1,N)$ of $F_t$ in \eqref{eq:TtQtSt} and block $(N,N)$ of $H$ in \eqref{eq:H_transformed} Cloud writes
	\bse
	\begin{align}
	2(A_t^N)^\top P_tB_t = (F_t)_{1,N} \label{eq:Fx0_R} \\
	2R + 2B_t^{\top}P_tB_t =H_{N,N}, \label{eq:Hx0_R}
	\end{align}
	\ese
	where $P_t \coloneqq T^{\top}PT$.
	From \eqref{eq:Fx0_R} Cloud obtains the \textit{unique} $P_tB_t$, and then from \eqref{eq:Hx0_R} it recovers unique matrix $R$.
	\mypar
 To recover $Q$ and $P$, Cloud needs to consider the equalities given in \eqref{eq:H_transformed} and \eqref{eq:TtQtSt} and form the set
	\be\label{eq:set_Qt_Pt_original}
	\begin{aligned}
		\Big\{  &Q_t, P_t \in \R^{n \times n}|{\calQ}_t  \coloneqq \bdiag({Q}_t, \ldots,{Q}_t ,{P}_t)\, \text{and}\\
		&\bbm \calS_t & \calT_t\ebm^\top {\calQ}_t \calS_t= \frac{1}{2}\col(H -2\calR, F_t) \Big\},
	\end{aligned}
	\ee
	where it knows the matrices $\calS_{t}$, $\calT_t$, $\calR$, $H$ and $F_t$. Note, the set \eqref{eq:set_Qt_Pt_original} is analogous to \eqref{eq:set_Q_P_original}, and therefore Cloud follows the same steps as in Theorem \ref{thrm:GOHF_known} to recover the set \eqref{eq:J_theorem_up_to_sim} for ${Q}_t$ and ${P}_t$. This gives the claims. $\qed$ 
\mypar
	\emph{Proof of Theorem \ref{thrm:HF_knwon}}. From Assumption \ref{Assum:sys_stable} it holds that $\lim_{N \to \infty} Y = \bY$.
	This enables Cloud to write using
	the block matrices $F_{1,1}$, $F_{1,2}$, $\ldots$, $F_{1,n+1}$ the following
	\be\label{eq:F_block_N_large}
	 \col(A^{\top}, \cdots, (A^{\top})^{n+1}) \bY B 
	 = \col(F_{1,1}, \cdots, F_{1,n+1}).
	\ee
From the above equation, it follows that
	\be\label{eq:A_eq_N_large}
	A^{\top} \underbrace{\bbm F_{1,1} & F_{1,2} & \ldots & F_{1,n} \ebm}_{\eqqcolon\calC} =  \bbm F_{1,2} & F_{1,3} & \ldots & F_{1,n+1} \ebm.
	\ee
	Observe that
	\begin{align*}
	\calC =A^\top \bbm  \bY B & (A)^{\top}\bY B & \cdots & (A^{n-1})^{\top}\bY B \ebm.		
	\end{align*}
By controllability of the pair $(A^\top, \bY B)$, equivalently observability of  $(A, B^\top \bY )$, and non-singularity of $A$, Cloud recovers the  state matrix $A$ from \eqref{eq:A_eq_N_large}.
For finding the matrix $B$, Cloud uses the known blocks $H_{2,1}$, $H_{3,1}$, $\ldots$, $H_{n+1,1}$ and writes
	\be\label{eq:B_eq_N_large}
	B^\top \calC  = 
	\bbm H_{2,1}^\top & H_{3,1}^\top & \cdots & H_{n+1,1}^\top\ebm.
	\ee
	From \eqref{eq:B_eq_N_large} and noting that $\calC$ has full row rank, Cloud finds $B$ uniquely. 
 Moreover, Cloud finds  $\bY B$ uniquely
	from \eqref{eq:F_block_N_large}.
	Then, from block $(1,1)$ of $H$, namely, 
	$
	R + B^{\top}\bY B = H_{1,1},
	$
	Cloud recovers the matrix $R$.
\mypar 
Now that Cloud has recovered the matrices $A$, $B$, and $R$, 
determining the matrices $Q$ and $P$ reduces to the same setup as in Theorem \ref{thrm:GOHF_known}, leading to the set in  \eqref{eq:J_theorem}. This completes the proof.$\qed$ 
\mypar 
\emph{Proof of Theorem \ref{thrm:Hx0F_knwon}}.
Cloud uses the block matrices $x_0^\top F_{1,1}$, $x_0^\top F_{1,2}$, $\ldots$, $x_0^\top F_{1,n}$ and writes:
	\be\label{eq:x_0F}
	x_0^\top \calC = x_0^\top \bbm A^\top \bY B & \cdots & (A^{\top})^{n}\bY B\ebm ,
	\ee
	where $\calC$ was defined in \eqref{eq:A_eq_N_large}.
	Recall the matrix of initial states defined in \eqref{eq:all_init} as
	$
	X_0 =\bbm x_{0|1} & x_{0|2} & \ldots & x_{0|I} \ebm.
	$
	Due to Assumption \ref{Assum: multiple_MPC}, Cloud writes \eqref{eq:x_0F} for $I$ instances as 
	\be\label{eq:all_init_cond}
	X_0^\top \calC = X_0^\top \bbm A^\top \bY B& \cdots & (A^{\top})^{n}\bY B\ebm .
	\ee
	Note, $\rank \calC =n$ because of the observability of the pair $(A, B^\top \bY )$ and invertibility of $A$.  Additionally, $\rank X_0 = n$ due to assumption. Hence, it follows $\rank X_0^\top \calC = n$. Then, Cloud writes a full factorization of $X_0^\top \calC$ as $X_0^\top \calC = X_b \calC_b$ where $X_b\in \R^{I \times n}$ and $\calC_b \in \R^{n \times nm}$ are full column rank and full row rank matrices. 
	Note that 
	\be\label{eq:X0_top_T}
	X_0^\top T = X_b, \quad T^{-1}\calC = \calC_b
	\ee
	for an unknown invertible $T \in \R^{n \times n}$ \cite[Prop. 7.6.6]{bernstein2018scalar}.
	 Analogous to \eqref{eq:all_init_cond}, Cloud uses the block matrices $x_0^\top F_{1,2}$, $x_0^\top F_{1,2}$, $\ldots$, $x_0^\top F_{1,n+1}$ and writes:
	\be\label{eq:At_X0}
		X_0^\top T T^{-1}A^\top \calC 
		 = X_0^\top \bbm (A^{\top})^{2} \bY B & \cdots & (A^{\top})^{n+1}\bY B\ebm.
	\ee
	By substituting $\calC = T\calC_b$ into \eqref{eq:At_X0}, Cloud recovers a unique value for $A_t \coloneqq T^\top A T^{-\top}$ (equivalently $T^{-1}A^\top T$) since it knows $X_0^\top T$ from \eqref{eq:X0_top_T} and $\calC_b$ (both matrices have full rank) and has received the right hand-side from Plant.  
\mypar
 For finding $B$ matrix,
	by considering the blocks $H_{2,1}$, $H_{3,1}$, $\ldots$, $H_{n+1,1}$ Cloud writes
	\be\label{eq:H matrices}
	B^\top T \calC_b  = 
	\bbm H_{2,1}^\top & H_{3,1}^\top & \cdots & H_{n+1,1}^\top\ebm.
	\ee
	Therefore, Cloud recovers unique value $T^\top B \eqqcolon B_t$ since $\calC_b$ is known. For recovering $R$, Cloud first rewrites \eqref{eq:H matrices} as 
	$$
	B^\top \bY T^{-\top} T^\top\bbm AB &  \cdots & A^nB\ebm = \bbm H_{2,1}  & \cdots & H_{n+1, 1}\ebm,
	$$
	and then obtains $B^\top \bY T^{-\top}$ since it knows
	  the full row rank matrix $T^\top\bbm AB & A^2B & \ldots & A^nB\ebm$.
	Then, it substitutes the value $(B^\top \bY T^{-\top})(T^\top B)$ into the block $H_{1,1}$, i.e.,
	$
	R + B^\top \bY B = H_{1,1}
	$,
	and recovers the unique matrix $R$. By replacing $T$ with $T^{-\top}$, we have the claims for the first part of Theorem. 	
\mypar
 For the second part of  Theorem, note that Cloud has recovered $A_t = T^{-1}AT$, $B_t = T^{-1}B$, and $R$ hence analogous to Theorem \ref{thrm:GWOx0HFx0} it can form the set given in \eqref{eq:set_Qt_Pt_original} and follow the subsequent steps to infer $Q_t$ and $P_t$. $\qed$ 
 \mypar 
\emph{Proof of Proposition \ref{prp:key_rev}}. Cloud receives $\tilde{G} = G\bR$. Recall $G$ in \eqref{eq:constraint_all_block_matrices}, i.e., $G =  \bbm I_{Nm}^\top & -I_{Nm}^\top & {\calG}^\top & -{\calG}^\top
	\ebm ^\top $. It is clear that Cloud recovers $\hat\bR = \bR$ from the results of the first matrix block in $\tilde{G}$. To recover $\br$, note  the rows $1$ to $Nm$ in $\tilde{e}$ read as
	\be\label{eq:umax_r}
	\begin{aligned}
		\bo_{N} \otimes u_{\text{max}} - \br &  \eqqcolon  \tilde{e}_1.
	\end{aligned}
	\ee
	Consider $L = I_{N} - (\frac{1}{N})\bo_{N}\bo_{N}^\top$ and multiply \eqref{eq:umax_r} from left by $L \otimes I_m$ to obtain $(L \otimes I_m)\br = -(L \otimes I_m) \tilde{e}_1$. Hence, Cloud recovers $\hat \br =(L \otimes I_m)\br$, which proves the claim. $\qed$ 
 \mypar
 \emph{Proof of Theorem \ref{thrm:iden_R_r_P_with_F}}. 
From the analysis preceding the theorem it follows that under Assumption \ref{Assum: multiple_MPC} and the condition $\rank F = n$, the  dynamical system \eqref{eq:delta_system} can be formed by Cloud, and its (known) data matrices satisfy \eqref{eq:tilde_e_system_Instances_F}. 
Furthermore, it can be observed from \eqref{eq:tilde_fi}  and \eqref{eq:delta_f_delta_zeta} that 
$\delta\tilde{f}_1, \cdots, \delta\tilde{f}_{I-1} \in  \im(\bR^\top F^{\top})$. Note also $\rank \bR^\top F^{\top} = n$ since $\bR$ is invertible and $\rank F = n$ due to assumption. By drawing on these observations, it follows that
\bse
\begin{align}
\bR^\top F^{\top} X_1 &=  \Delta\tilde{F}_0 \label{eq:RtFtX1} \\
\bR^\top F^{\top} X_2 &=  \Delta\tilde{F}_1,\label{eq:deltaF1}
\end{align}
\ese
where $X_1 \in \R^{n \times (I-2)}$ is full row rank, $ X_2 \in \R^{n \times (I-2)}$ and both are unknown to Cloud since $\bR^\top F^{\top}$ is unknown. Substituting $ \Delta\tilde{F}_0$ from \eqref{eq:E0_data} in \eqref{eq:RtFtX1} and $\bR^\top F^{\top}$ from \eqref{eq:RtFtX1} in \eqref{eq:deltaF1}  yields 
\be\label{eq:E1_data}
\begin{aligned}
	\bR^\top F^{\top} &= {\Delta F_b} T\\
		\Delta\tilde{F}_1 &= {\Delta F_b} E_1 
\end{aligned}
\ee
for Cloud where $E_0X_1^\dagger \eqqcolon T \in \R^{n \times n}$ is  invertible and $E_0X_1^\dagger X_2 \eqqcolon E_1 \in \R^{n \times (I-2)}$. Note, while Cloud knows $E_1$, it does not know $T$.
By substituting $\Delta\tilde{F}_0$ given in \eqref{eq:E0_data}, $\Delta\tilde{F}_1$ and $\bR^\top F^{\top}$ given in \eqref{eq:E1_data} into \eqref{eq:tilde_e_system_Instances_F} it follows that 
$$
E_1 = {\Delta F_b}^\dagger\tilde{A}{\Delta F_b} E_0 + {\Delta F_b}^\dagger\tilde{B}\Delta\tilde{Z}^*,
$$
which is further simplified as
\be\label{eq:system_data_final}
E_1 = \bbm TAT^{-1} &  T\bar{B}\bR\ebm \bbm E_0 \\ \Delta\tilde{Z}^* \ebm.
\ee
Multiplying both sides
of the above equality from the right by $\Theta$  satisfying \eqref{eq:rank_cond_data} yields $E_1 \Theta = TAT^{-1}$, and thus Cloud identifies 
$\hat{A} = TAT^{-1}$ as claimed. $\qed$ 
 
\section{Dense MPC matrices \eqref{eq:LQ_merged}}\label{app:HFY}
 We consider the matrices in \eqref{eq:LQ_merged} to be block partitioned based on $N$, and refer to an arbitrary block by mentioning its position in the matrix. For instance, we partition $F$ in  the cost function as $F = \bbm F_{i,j}\ebm$ where the block $F_{i,j} \in \R^{n \times m}$ and $i = 1$, $j = 1, 2, \ldots, N$.
\mypar 
The matrices $H$, $F$, and $Y$ in \eqref{eq:cost_LQ_merged} can be expanded as:
\small
\begin{align*}
\frac{1}{2}H = \calR + {\calS}^{\top}{\calQ}{\calS}  =  \bbm H_{1,1} & \star & \star & \ldots & \star \\
H_{2,1} & H_{2,2} & \star & \ldots & \star \\
\vdots & \vdots & \ddots & \vdots & \vdots \\
H_{N-1,1} & H_{N-1,2} &  & H_{N-1,N-1} & \star \\
H_{N,1} & H_{N,2} & \ldots & H_{N,N-1} & H_{N,N}
\ebm
\end{align*}
\begin{align*}
& H_{1,1} = R + B^\top\big(\sum_{i=0}^{N-2}{(A^{\top})^{i}QA^{i}} + (A^{\top})^{N-1}PA^{N-1}\big)B\\
& H_{2,1} = B^\top\big(\sum_{i=0}^{N-3}{(A^{\top})^{i}QA^{i}} + (A^{\top})^{N-2}PA^{N-2}\big)AB \\
& \qquad \qquad \vdots \\
& H_{N-1,1} = B^\top \big(Q + A^{\top}PA\big)A^{N-2}B\\
&	H_{N,1} = B^\top PA^{N-1}B
\end{align*}
\begin{align*}
&H_{2,2} = R + B^\top\big(\sum_{i=0}^{N-3}{(A^{\top})^{i}QA^{i}} + (A^{\top})^{N-2}PA^{N-2}\big)B\\
& \qquad \qquad \vdots \\
& H_{N-1,2} = B^\top \big(Q + A^{\top}PA\big)A^{N-3}B\\
&H_{N,2} = B^\top PA^{N-2}B\\
& H_{N-1,N-1} = R + B^\top\big(Q + A^{\top}PA\big)B \\
&H_{N,N-1} = B^\top PAB\\
&	H_{N,N} = R + B^\top PB
\end{align*}
\begin{align*}
\frac{1}{2}F =& {\calT}^{\top}{\calQ}{\calS} =  \bbm 
F_{1,1} & F_{1,2} & \ldots & F_{1,N-1} & F_{1,N}
\ebm \\
&F_{1,1} = A^\top\big(\sum_{i=0}^{N-2}{(A^{\top})^{i}QA^{i}} + (A^{\top})^{N-1}PA^{N-1}\big)B \\
& F_{1,2} = (A^2)^\top\big(\sum_{i=0}^{N-3}{(A^{\top})^{i}QA^{i}} + (A^{\top})^{N-2}PA^{N-2}\big)B 
\end{align*}
\begin{align*}
& \qquad \qquad \vdots \\
&F_{1,N-2} = (A^{N-2})^\top\big(Q + A^{\top}QA + (A^{2})^{\top}PA^{2})B\\
&F_{1,N-1} = (A^{N-1})^\top\big(Q + A^{\top}PA)B \\
& F_{1,N} = (A^{N})^\top PB
\end{align*}
\begin{align*}
Y = Q+{\calT}^\top{\calQ}{\calT} = \sum_{i=0}^{N-1}{(A^{\top})^{i}QA^{i}} + (A^{\top})^{N}PA^{N}
\end{align*}
\normalsize

\bibliography{./MyReferences}

\end{document}